\documentclass[aps,prl,twocolumn,superscriptaddress,hidelinks,nobibnotes,nofootinbib]{revtex4-2}
\usepackage{graphicx}
\usepackage{physics}
\usepackage{amssymb}
\usepackage{hyperref}
\usepackage{xcolor}
\usepackage{soul}
\usepackage{bm}
\usepackage{mathtools}
\usepackage[symbol, hang]{footmisc}

\hypersetup{
    colorlinks,
    linkcolor=[RGB]{46, 48, 146},
    citecolor=[RGB]{46, 48, 146},
    urlcolor=[RGB]{46, 48, 146}
}

\begin{document}
\footnotetext{These authors contributed equally to this work.}
\footnotetext{nathan.lysne@quantinuum.com}
\footnotetext{Present Address: Quantinuum K.K., Tokyo, Japan.}
\footnotetext{dietrich.leibfried@nist.gov}

\title{Individual addressing and state readout of trapped ions utilizing rf micromotion}

\author{Nathan K. Lysne}

\affiliation{Time and Frequency Division, National Institute of Standards and Technology, 325 Broadway, Boulder, Colorado 80305, USA}
\affiliation{Department of Physics, University of Colorado, Boulder, Colorado 80309, USA}

\author{$\!\!^{,*,\dagger,\ddag}$\ Justin F. Niedermeyer}
\affiliation{Time and Frequency Division, National Institute of Standards and Technology, 325 Broadway, Boulder, Colorado 80305, USA}
\affiliation{Department of Physics, University of Colorado, Boulder, Colorado 80309, USA}
\author{$\!\! ^{,*}$\ Andrew C. Wilson}
\affiliation{Time and Frequency Division, National Institute of Standards and Technology, 325 Broadway, Boulder, Colorado 80305, USA}
\author{Daniel H. Slichter}
\affiliation{Time and Frequency Division, National Institute of Standards and Technology, 325 Broadway, Boulder, Colorado 80305, USA}
\author{Dietrich Leibfried$^{1,\text{\textsection}}$}
\noaffiliation

\date{July 16, 2024}

\begin{abstract}
\noindent Excess ``micromotion'' of trapped ions due to the residual radio-frequency (rf) trapping field at their location is often undesirable and is usually carefully minimized. Here, we induce precise amounts of excess micromotion on individual ions by adjusting the local static electric field they experience. Micromotion modulates the coupling of an ion to laser fields, ideally tuning it from its maximum value to zero as the ion is moved away from the trap's rf null. We use tunable micromotion to vary the Rabi frequency of stimulated Raman transitions over two orders of magnitude, and to individually control the rates of resonant fluorescence from three ions under global laser illumination without any changes to the driving light fields. The technique is amenable to situations where addressing individual ions with focused laser beams is challenging, such as tightly packed linear ion strings or two-dimensional ion arrays illuminated from the side.
\end{abstract}

\maketitle

Trapped ions are used for a wide range of quantum applications, including precise time-keeping \cite{ludlow_atomicclocks_2015, huntemann_clock_2016,brewer_prl_2019,burt_space_2021}, quantum simulation \cite{blatt_naturep_2012,hempel_qchem_2018,monroe_rmp_2021,shapira_timereversal_2023}, quantum information processing \cite{blatt_nature_2008, Bruzewicz2019,erhard_latsurgery_2021,egan_ftc_2021,ryananderson_qec_2022,postler_ftuqgate_2022} and tests of fundamental physical theories \cite{godun_timevar_2014,huntemann_massvar_2014,pruttivarasin_nature_2015,dreissen_lorentz_2022,kozlov_hci_2018,arrowsmith_radmol_2023}. In these applications, individually-addressed single- and two-qubit operations are often realized with tightly-focused laser beams \cite{wang_twoqubit_MEMS_2020,kranzel_strings_2022,joshi_hydrodynamics_2022}, sometimes in combination with transport of ions into and out of trap regions with laser illumination \cite{wineland_experimental_1998, kielpinski_large_scale_2002, wan_teleport_zones_2019, pino_ccd_demonstration_2021,moses_race_2023}. Performing precise, individually addressed operations in this manner for ions trapped in linear arrangements 
requires a growing overhead in the number of beams or the number of transport operations as the number of ions increases. The task becomes even more demanding if the ions are held in a two-dimensional (2D) array that is illuminated from the side \cite{wineland_experimental_1998,kiesenhofer_2dions_2023,moses_race_2023}, because of the increased potential for spurious illumination of ions at different positions along the laser beam.

Techniques for individual addressing not mediated by tightly focused laser beams rely on the use of additional spatially-varying electric or magnetic fields to resolve the addressed ions in position or frequency \cite{wineland_experimental_1998, turchette_ion_entanglement_1998, leibfried_individual_1999, mintert_long_wave_logic_2001, staanum2002, chiaverini_qecc_2004, mchugh2005, haljan2005, chiaverini2008, johanning2009, wang2009, warring_mw_gradient_2013, navon_variable_coupling_2013, piltz_q_register_2014, aude-craik_microwave_singlesite_addr_2014, seck_mod_potential_addressing_2020, srinivas_high_fid_laser_free_2021, sutherland_geo_phase_addressing_2023, srinivas_local_fields_2023}. One such approach for individual addressing exploits precise control of the driven periodic motion of the ion at the frequency of the rf trapping field, known as micromotion \cite{berkeland_minimization_1998, turchette_ion_entanglement_1998, leibfried_individual_1999, navon_variable_coupling_2013}. Micromotion is often viewed as undesirable, because it can interfere with the ability to control trapped ions and gives rise to undesired systematic shifts for trapped ion frequency standards \cite{berkeland_minimization_1998, brewer_prl_2019}. As a result, various techniques to compensate for stray electric fields and to move ions closer to a micromotion minimum have been developed and are routinely used \cite{berkeland_minimization_1998, keller_precise_2015,nadlinger_micromotion_2021}. 

However, controlled reintroduction of excess micromotion can be a resource for individually controlling the coupling of ions to light fields, especially as the scale of trapped ion systems grows \cite{turchette_ion_entanglement_1998,leibfried_individual_1999,navon_variable_coupling_2013}. Recently there has been renewed interest in the controlled introduction of micromotion to reduce quantum logic gate errors \cite{bermudez_micromotion-enabled_2017} and to suppress decoherence of bystander ions during mid-circuit fluorescence readout \cite{gaebler_suppression_2021}. 

In this work, we induce precise amounts of excess micromotion along the wave vector of an applied laser light field to selectively change the coupling of the ion to the light. In the reference frame of the ion, the light appears frequency modulated, and the light intensity is spectrally redistributed from the rest-frame carrier frequency to sidebands detuned by multiples of the trap rf drive frequency (``micromotion sidebands'') \cite{wineland_laser_cooling_1979, wineland_experimental_1998}. Building on previous work that used excess micromotion for spectrally resolved addressing or suppressing localized gate errors, we expand this technique for individual addressing and readout of a larger number of ions using unmodulated globally applied control fields. As a proof of principle, we demonstrate tuning the Rabi frequency of a stimulated Raman transition over two orders of magnitude, achieving a suppression of the Rabi rate by a factor of up to $110.8 \pm 1.2$ (68\% confidence). We also demonstrate individual fluorescence readout of three ions under global laser illumination by selectively tuning their interactions with the laser field, which is resonant with a cycling transition of the ions.

Micromotion is periodic motion of a trapped ion driven by the rf trapping field. While a certain amount of micromotion is unavoidable \cite{wineland_experimental_1998,keller_precise_2015}, excess micromotion can be induced by displacing the ion from the rf null of the trap. We consider a total potential for the ion $\Phi$ that is a sum of an effective rf pseudopotential $\Phi_{\text{pp}}$ \cite{dehmelt_rf_spectra_i_1968,douglas_effective_rf_motion_2015} and the static electric potential $\Phi_{\text{dc}}$ (including stray fields).

By applying a static electric field $\textbf{E}_{\text{shift}}$, the ion will be displaced from its original equilibrium position such that the force from $\textbf{E}_{\text{shift}}$ on the ion is balanced by the restoring force due to the potential gradient
\begin{equation}\label{eq_equilibrium}
q \mathbf{E}_{\text{shift}}(\mathbf{r}_e) = -q \nabla \Phi(\mathbf{r}_e)\, ,
\end{equation}
\noindent where $q$ is the charge of the ion and $\mathbf{r}_e$ is the new equilibrium position. The ion undergoes driven oscillations at the rf trapping frequency $\Omega_{\text{rf}}/2\pi$ of amplitude $\delta \mathbf{r}(t)$ around $\mathbf{r}_e$, described in the approximation of uniform rf electric field over the spatial extent of the ion oscillations by
\begin{equation}
\delta \mathbf{r}(t) = \frac{q \mathbf{E_{\text{rf}}(\mathbf{r}_{\textit{e}})}}{m \Omega_{\text{rf}}^2}\cos(\Omega_{\text{rf}} t).
\end{equation}
This is equivalent to the forced motion of a free particle of charge $q$ and mass $m$ due to the rf electric field $\mathbf{E_{\text{rf}}}$ at $\mathbf{r}_e$. The direction of the micromotion $\delta \mathbf{r}(t)$ depends on the trap design and is not in general parallel to $\textbf{E}_{\text{shift}}$.

A propagating electric field $\mathbf{E}_\ell$ from a laser beam with wave vector $\mathbf{k}$, frequency $\omega_\ell$, and phase $\phi_\ell$ appears modulated in the frame of the shifted, oscillating ion as
\begin{eqnarray}\label{eq_sidebands}
\mathbf{E}_\ell\left(\mathbf{r}(t),t\right) \approx\mathbf{E}_\text{0}(\mathbf{r}_e)   \exp[i(\mathbf{k} {\cdot} \{\mathbf{r}_e {+}\delta \mathbf{r}(t)\}{-} \omega_\ell t {+}\phi_\ell)] \nonumber \\
\noindent = \mathbf{E}_\text{0}(\mathbf{r}_e) e^{i(\mathbf{k} \cdot\mathbf{r}_e+\phi_\ell)}\sum_{\mathclap{n=-\infty}}^\infty J_n(\beta)\exp[i (n \Omega_{\text{rf}}{-}\omega_\ell)t],
\end{eqnarray}

where we have used the Jacobi-Anger identity with $J_n$ as the $n$th Bessel function of the first kind and the modulation index
\begin{equation}\label{eq_modind}
\beta =\frac{q}{m \Omega^2_{\text{rf}}} \mathbf{k} \cdot\mathbf{E}_{\text{rf}}(\mathbf{r}_e).
\end{equation}
A suitable static field $\mathbf{E}_{\text{shift}}$ induces micromotion with a component along $\mathbf{k}$ and scales the component of the electric field at the carrier frequency $(n=0)$ by $J_{0}(\beta)$, thereby tuning the strength of the interaction between an ion and incident resonant light without changing the light field itself \cite{turchette_ion_entanglement_1998,leibfried_individual_1999}. For the case when $\beta_{j}$ is the $j$th Bessel zero, $J_0(\beta_{j}) = 0$ (e.g. $\beta_{1} \approx 2.4048$), the component of the field at the carrier frequency is completely suppressed and all of the power of the light field is shifted to the first and higher-order sidebands ($|n| \geq1)$. Similar considerations hold for stimulated Raman transitions driven by two light fields with frequencies $\omega_1, \omega_2$ and wave vectors $\mathbf{k}_1, \mathbf{k}_2$ after replacing $\mathbf{k} \rightarrow \mathbf{k}_2-\mathbf{k}_1 \equiv \mathbf{\Delta k}_{\text{Ra}}$ and $\omega_{\ell} \rightarrow \omega_2-\omega_1$. For Raman transitions between closely spaced states with $|\omega_2-\omega_1|\ll \{\omega_2,\omega_1\}$, it is advantageous to use a ``motion-sensitive'' configuration with non-copropagating light fields to increase $\mathbf{\Delta k}_{\text{Ra}}$.

\begin{figure}[t]
\includegraphics[width=8.4cm]{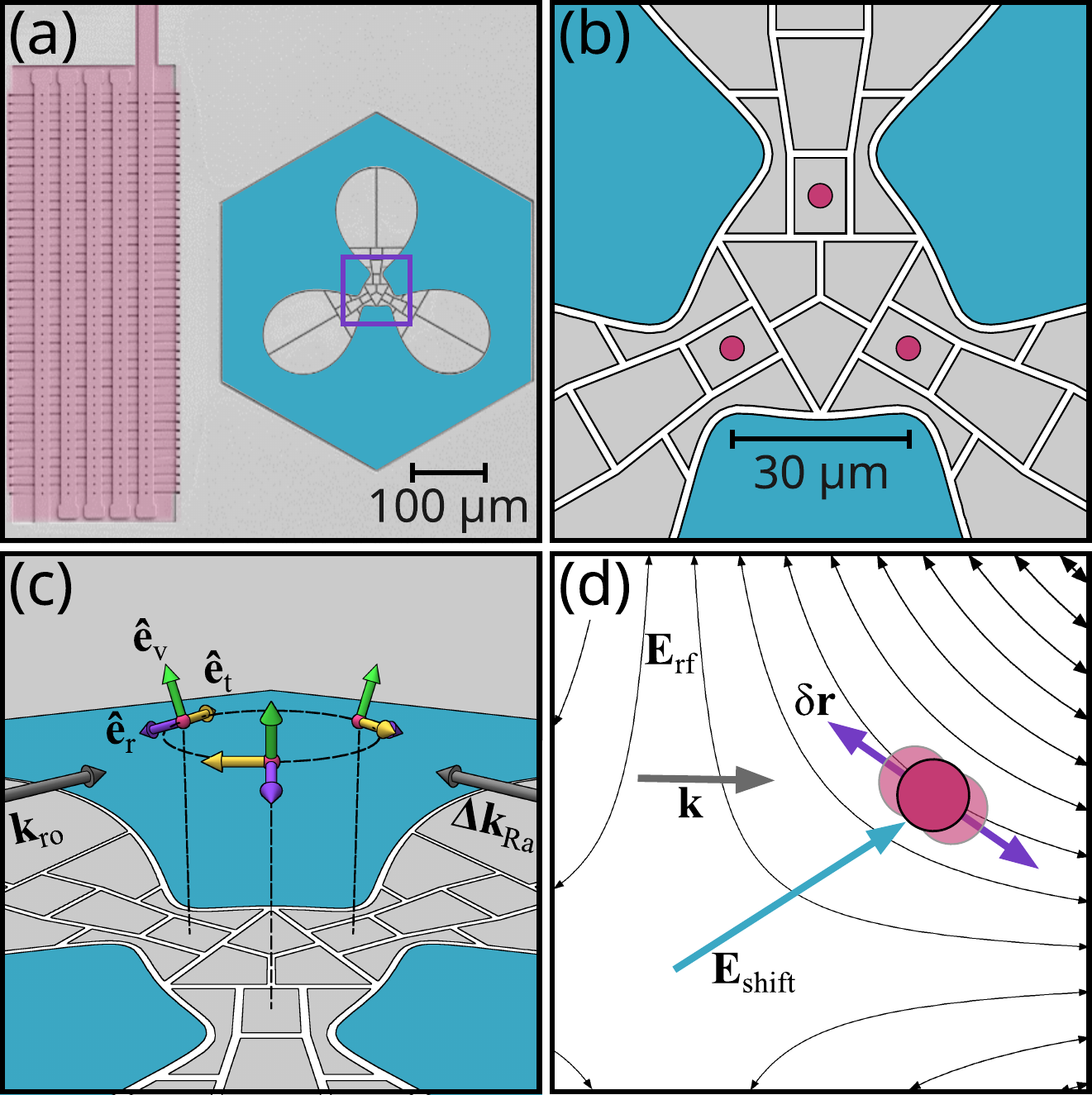}
\caption{Trap configuration and principle of micromotion addressing. (a) Top-view false-color scanning electron micrograph of the trap. The rf electrode (blue) surrounds an array of 30 control electrodes (gray) to which static potentials can be applied. An integrated microwave antenna (red) for driving hyperfine transitions is at left. (b) Detail of the purple square region in (a), with red dots denoting the positions of the three trapping sites. (c) Perspective view of the region in (b), with arrows indicating the directions of the radial (purple), vertical (green), and tangential (yellow) motional modes ($\mathbf{\hat{e}}_r$, $\mathbf{\hat{e}}_v$, and $\mathbf{\hat{e}}_t$, respectively) for each site. The radial and vertical modes are tilted ${\approx}19.5^\circ$ from the electrode plane and the direction normal to the plane, respectively, while the tangential modes are tangent to the circle connecting all three sites. The mode directions have $120^\circ$ rotational symmetry about the normal to the electrode plane at the center of the triangle. The wave vector $\mathbf{k}_{\text{ro}}$ of the readout laser beam and the difference wave vector $\mathbf{\Delta k}_{\text{Ra}}$ of the Raman laser beams are shown as black arrows. (d) Schematic drawing of an ion undergoing micromotion when displaced from the rf null by an applied field $\textbf{E}_{\text{shift}}$. Excess micromotion induced by $\textbf{E}_{\text{shift}}$ modulates the coupling between the ion and an incident laser light field proportional to the overlap of the wave vector $\mathbf{k}$ and the micromotion vector $\mathbf{\delta r}$.}
\label{fig_ttrap}
\end{figure}

Experiments are performed in a surface electrode trap designed at NIST and fabricated by Sandia National Laboratories [see Fig.\,\ref{fig_ttrap}]. A single electrode driven at $\Omega_{\text{rf}}/{2 \pi} \approx 121.1 \text{ MHz}$ produces zeros of $\mathbf{E}_{\text{rf}}$ in three locations (sites) arranged in an equilateral triangle with 30 \textmu m side length at a distance of 40 \textmu m from the electrode plane. A single $^{9}\text{Be}^{+}$ ion trapped in one of these sites has three orthogonal modes of motion [see Fig.\,\ref{fig_ttrap}(c) for mode directions]. With an amplitude $U_{\text{rf}}\approx 44 \text{ V}$ applied to the rf electrode, the motional mode frequencies are $\left\{\omega_\text{r},\omega_\text{v},\omega_\text{t}\right\}/ {2 \pi} \approx \left\{3.9, 7.0, 10.9\right\} \text{ MHz}$, for the radial, vertical and tangential modes respectively. 
The corresponding motional heating rates were ${\approx}
 \left\{30,15,2\right\} \text{ quanta/s}$ respectively for all sites when operating at $5.5 \text{ K}$.

A static $0.5 \text{ mT}$ field lifts the energy degeneracy of hyperfine states in the $^2S_{1/2}$ electronic ground state manifold and defines a quantization axis. We optically pump to the state $\left|F{=}2,m_{F}{=}-2\right\rangle\equiv\left|\downarrow\right\rangle$ to prepare the internal state of the ions, and then couple to the other qubit state $\left|1,-1\right\rangle\equiv\left|\uparrow\right\rangle$ using microwave or stimulated Raman transitions \cite{wineland_experimental_1998}. The splitting between these states is ${\sim} 1.26 \text{ GHz}$. Other hyperfine transitions within the $^2S_{1/2}$ manifold can be driven with lasers or with microwave tones applied to the trap-integrated microwave antenna (see Fig.~\ref{fig_ttrap}(a)). A laser beam at $313 \text{ nm}$ is used for Doppler cooling and state-dependent fluorescence readout \cite{wineland_experimental_1998} on the closed $\left|\downarrow\right\rangle\leftrightarrow {^2P_{3/2}} \left|3,-3\right\rangle$ cycling transition. The beam is directed parallel to the electrode plane along the quantization axis [$\mathbf{k}_{\text{ro}}$ shown in Fig.\,\ref{fig_ttrap}(c)], and has an elliptical cross section with a 10:1 aspect ratio such that it illuminates the three ions nearly equally. Ion fluorescence is collected with an objective lens and imaged onto a camera or photomultiplier tube (PMT). Two counter-propagating beams (difference wave vector $\mathbf{\Delta k}_{\text{Ra}}$) perpendicular to the cooling and detection beam are focused such that they illuminate only a single site. These beams are detuned $\approx\!80 \text{ GHz}$ blue from the $^2S_{1/2}\leftrightarrow^2P_{1/2}$ transition and are used to drive stimulated Raman transitions. 

\begin{figure}
\includegraphics[width=8.4cm]{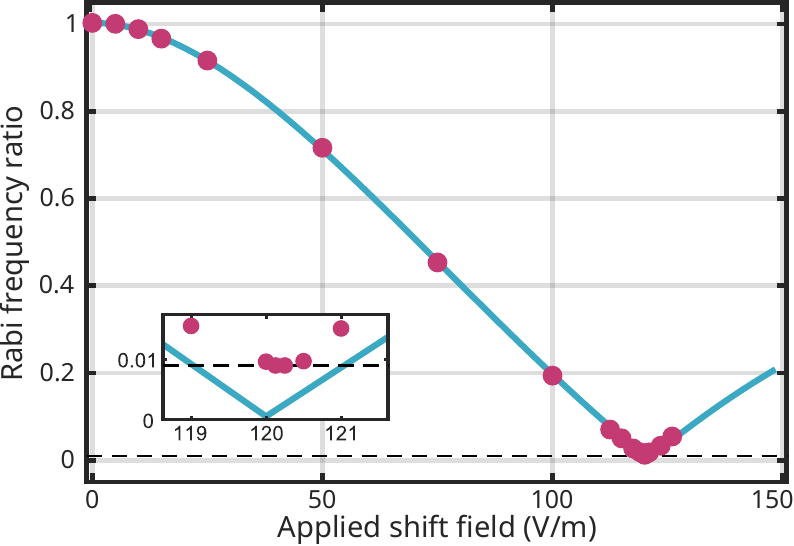}
\caption{Dependence of the Raman carrier Rabi frequency on the applied shift field $\mathbf{E}_{\text{shift}}$. The ratio $\Omega_\text{R}/\Omega_\text{R0}$ is shown as a function of the electric field applied along the radial direction. The experimentally determined ratios (red disks) agree with a theoretical model (blue line) described in the text. The minimum measured value of this ratio (dotted line) corresponds to a suppression of $\Omega_\text{R}$ by over two orders of magnitude. The inset shows a magnified view centered around the theoretical minimum of the ratio. Error bars at 68\% confidence for each point are smaller than the plotted markers.}
\label{fig_rabiShift}
\end{figure}

\begin{figure}
\includegraphics[width=8.4cm]{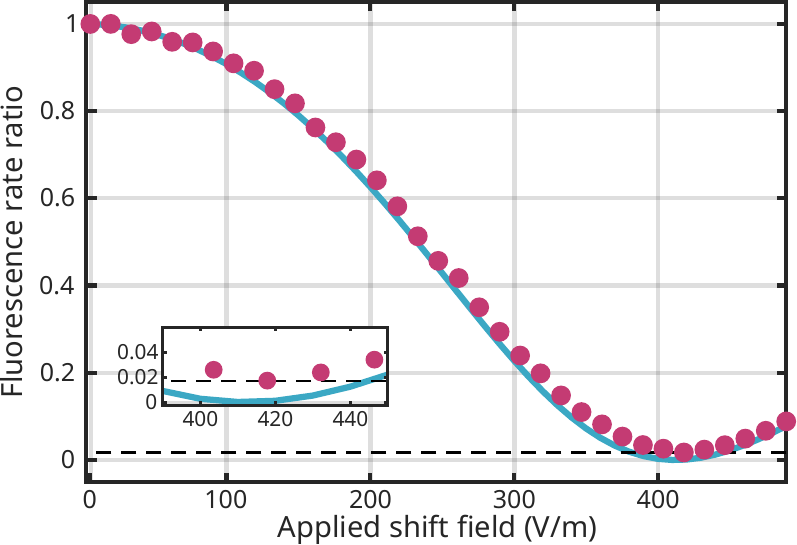}
\caption{Resonant fluorescence suppression from induced micromotion. The normalized fluorescence rate (relative to that at zero shifting field) of an ion prepared into the bright state (red disks) is shown versus the shifting field applied, and compared to the theoretical model described in the text (blue line). The inset shows a magnified view centered around the shift field needed to reach $\beta_{1}$, highlighting the minimum fluorescence measured (dotted line). Error bars at 68\% confidence for each point are smaller than the plotted markers.}
\label{fig_individualShift}
\end{figure}

Electric fields and potential curvatures (components of the tensor $\nabla^2\Phi$) in all three sites can be altered independently by applying suitable linear combinations of potentials to the 30 control electrodes. In this way it is possible to produce electric fields of arbitrary spatial orientation, as well as arbitrary curvatures, in a certain site without altering fields or curvatures in the other two sites. These ``shim'' fields from the control electrodes can be used to compensate for stray fields so that the ion equilibrium position in each site is at the rf null. Likewise, stray curvatures can be compensated or motional mode frequencies and directions can be tuned in individual sites by other suitable applied potentials.

After compensation, excess micromotion can be induced by generating an additional field $\mathbf{E}_{\text{shift}}$ with the same electrodes to move an ion to a new equilibrium point given by Eq.~(\ref{eq_equilibrium}). If the excess micromotion produced in this way has a component along the wave vector of a light field, the light will be modulated in the ion frame according to Eq.~(\ref{eq_sidebands}). We use a gapless Biot-Savart type model \cite{wesenberg_electrostatics_2008} of the trap, combined with the experimentally measured secular frequencies at the rf nulls, to calculate $\Phi_\text{pp}$, $\Phi_\text{dc}$, $\mathbf{E}_\text{shift}$, and $\mathbf{E}_\text{rf}$ given the applied voltages. From Eqs.~(\ref{eq_equilibrium})-(\ref{eq_modind}) we can then calculate $\mathbf{r}_e$ and $\beta$ versus $\mathbf{E}_\text{shift}$. For example, we calculate that the ion in the lower right site in Fig.\,\ref{fig_ttrap}{}(b) can be pushed $\approx\!2.37 \text{ \textmu m}$ along the radial mode direction by a field of magnitude $|\mathbf{E}_{\text{shift}}|\approx 208 \text{ V/m}$. In the new equilibrium position $\mathbf{r}_e$,  $|\mathbf{E}_{\text{rf}}(\mathbf{r}_e)|\approx 6980 \text{ V/m}$ and the micromotion makes an angle of approximately 22.5$^\text{o}$ with the wave vector of the detection beam, yielding $\beta\!\approx\!\beta_{1}$ with a micromotion amplitude of $\approx\!130$~nm. We checked for micromotion-related heating by measuring the radial mode heating rates in one site, which were statistically indistinguishable for $|\mathbf{E}_\text{shift}|= 0 \text{ V/m}$ and $213 \text{ V/m}$.

\begin{figure*}[]
\includegraphics[width=\textwidth]{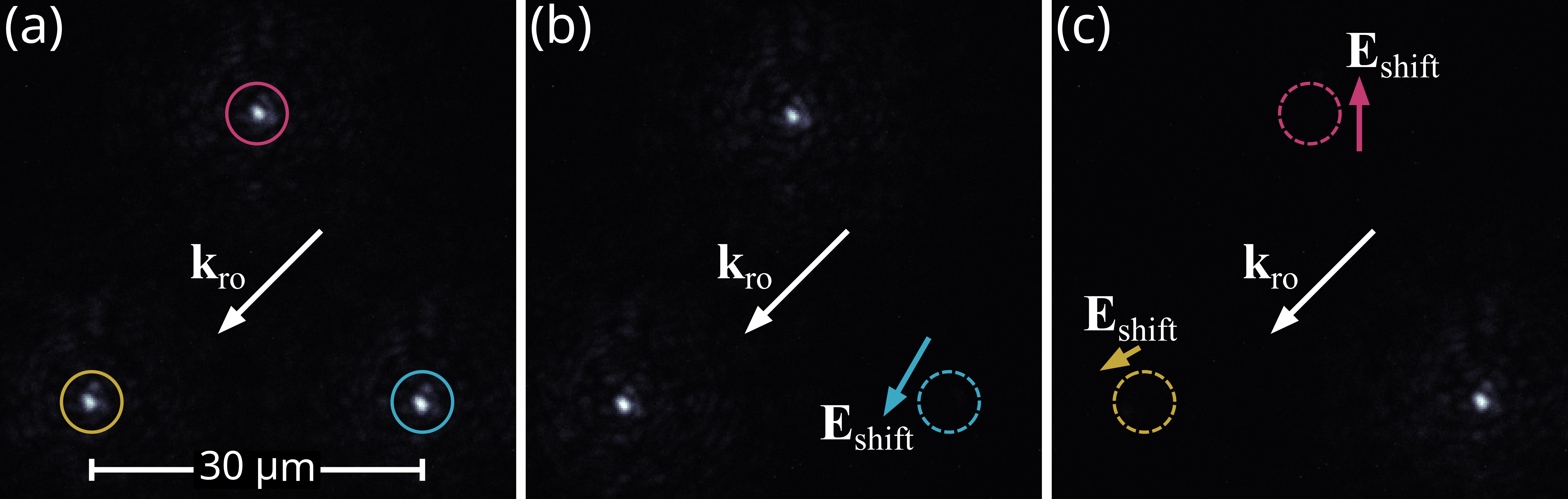}
\caption{Camera images of selectively shifted ions. In all images all ions are prepared in a bright state $\left|\downarrow\right\rangle$ and uniformly illuminated. (a) Three ions illuminated in the triangular array. In this case, no shifting field is applied. (b) A suitable shift (colored arrow) renders a single ion nearly invisible. (c) Any of the ions can be read out individually by shifting the other two. Nearly identical results were observed for all permutations of shifting individual and pairs of ions. Shift directions and relative strengths are indicated by the corresponding vectors.}
\label{fig_shiftPhotos}
\end{figure*}

 The limit of suppression is set by off-resonant scattering on the micromotion sidebands, as described in Eq.(1) in Ref. \cite{gaebler_suppression_2021}. For a narrow Raman transition where the Rabi rate for an unshifted ion $\Omega_\text{R0} \lll \Omega_{\text{rf}}$, all micromotion sidebands are far off-resonant, and the carrier coupling can be tuned almost to zero by induced micromotion. To show this experimentally we perform Doppler and pulsed sideband cooling of all three modes of an ion in the leftmost site in Fig.\,\ref{fig_ttrap}(b) to near its ground state. We change $\mathbf{E}_{\text{shift}}$ by a certain amount and drive the Raman carrier transition $\left|\downarrow\right\rangle \leftrightarrow \left|\uparrow\right\rangle$ for a variable duration. We then return $\mathbf{E}_{\text{shift}}$ to zero and detect the population in $\left|\downarrow\right\rangle$. We fit the resulting Rabi oscillations to find the Rabi frequency $\Omega_\text{R}$ as a function of $\mathbf{E}_{\text{shift}}$. 
 
Fig.\,\ref{fig_rabiShift} plots the ratio $\Omega_\text{R}/\Omega_\text{R0}$ as a function of the applied shift field, where $\Omega_\text{R0}/{2\pi} = 385.6 \pm 1.9 \text{ kHz}$ is the Rabi rate at the rf null. The ratio reaches a sharp minimum of $\Omega_\text{R}/\Omega_\text{R0}=(9.0 \pm 0.1)\times 10^{-3}$, or equivalently a carrier Rabi rate suppression factor of $110.8 \pm 1.2$, at calculated values of $|\mathbf{E}_\text{shift}|= 120 \text{ V/m}$ and $\mathbf{r}_e = 1.18 \text{ \textmu m}$. The angle between $\mathbf{\Delta k}_{\text{Ra}}$ and the rf electric field with calculated modulus $|\mathbf{E}_\text{rf}(\mathbf{r}_e)|= 3350 \text{ V/m}$ is 15.3$^\text{o}$. As a larger shift field is applied, the component of the addressing field at the carrier frequency, and thus the coupling between the ion and Raman beams, increases again as anticipated.

We compare these results with theoretical predictions based on the calculated micromotion amplitude and direction at the shifted ion position. The resulting curve, with no free parameters, is shown in blue in Fig.~\ref{fig_rabiShift} and agrees well with the measured data. We attribute the discrepancy between theory and experiment at the highest Rabi rate suppression to small fluctuations in the rf and static voltages applied to the trap electrodes. 

Resonant scattering, such as on a cycling transition in fluorescence readout, can also be suppressed by inducing micromotion \cite{leibfried_individual_1999,gaebler_suppression_2021}. The relatively broader linewidth of this transition compared to $\Omega_\text{rf}$ limits the ratio of the minimum scattering rate to the maximum scattering rate to $4.3 {\times} 10^{-3}$ \cite{gaebler_suppression_2021}. To demonstrate fluorescence suppression experimentally, we Doppler cool the ion and optically pump it into $\left|\downarrow\right\rangle$. We then apply $\mathbf{E}_{\text{shift}}$ and illuminate the ion with resonant light on the cycling transition. With no shifting field applied we detect an average of 10.7 counts on the PMT within $500 \text{ \textmu s}$. The ratio of count rate when applying $\mathbf{E}_{\text{shift}}$ to the field-free count rate is shown in Fig.~\ref{fig_individualShift}. 
The normalized fluorescence rate ratio as a function of shift field is in close agreement with the theoretical model based on the carrier scattering rate (proportional to $J_0(\beta)^2$ \cite{berkeland_minimization_1998}) with effects due to saturation of the cycling transition included. Besides the limit imposed by off-resonant scattering on the micromotion sidebands, the observed minimal ratio is affected by the detector dark counts, stray detection light, and the degree of saturation of the transition with no shift applied. Still, the minimal scattering rate of the ion is nearly indistinguishable from that of an ion that does not participate in a cycling transition, or from the case of no ion in the trap. 

Individual readout of multiple ions is possible without individual optical addressing if shifting fields can be selectively applied to any subset of ions. We experimentally demonstrate this by trapping one ion in each of the three sites, preparing all ions in $\left|\downarrow\right\rangle$, and illuminating them nearly uniformly with the elliptical beam that resonantly drives the cycling transition.

After minimizing excess micromotion, all three ions are nearly uniformly bright in an electron-multiplying charge-coupled device (EMCCD) camera image with a $15 \text{ s}$ exposure, shown in Fig.\,\ref{fig_shiftPhotos}(a). By displacing one of the ions to where it minimally scatters, it can be verified that the combinations of potentials used in each case have negligible effect on the scattering rate of the ions in the two other sites [Fig.\,\ref{fig_shiftPhotos}(b)]. Individual readout requires displacing two ions at the same time, as shown in Fig.\,\ref{fig_shiftPhotos}(c). We independently verified that the count histograms with two displaced and one bright ion are consistent with the histogram of a single bright ion trapped in any of the three sites. When all three ions are displaced simultaneously, an independently measured histogram is nearly indistinguishable from one taken with all three ions prepared in the dark $\left|\uparrow\right\rangle$ state and ``shelved'' to the $\left|1,1\right\rangle$ state with microwave pulses. These results certify that this technique is suitable for individual readout of three ions in this trap with no substantial loss in fidelity.

In summary, we have demonstrated the use of well-controlled shift fields to induce excess micromotion, thus selectively tuning the coupling of ions to incident light fields without modulating the driving fields themselves. We tune a Rabi frequency over two orders of magnitude, currently limited by small fluctuations of the static and rf potentials of the trap. We have also demonstrated readout of a single ion in an array by selectively suppressing the scattering of resonant readout light by neighboring ions. The degree of control demonstrated in this work compares favorably with other recent results in multi-site traps \cite{gaebler_suppression_2021,kwon_multisite_optical_2023,srinivas_local_fields_2023}. While the particulars of the implementation will vary, this technique can be implemented with any trap capable of compensating for micromotion by applying electric potentials to electrodes \cite{leibfried_individual_1999}. Independent, locally tunable coupling to global laser beams can also be used to drive quantum gates simultaneously in multiple array sites, reducing time overhead. Micromotion addressing can also be used in conjunction with other techniques to achieve finer control over coupling to specific ions, such as integrated optical waveguides \cite{mehta_inegrated_optics_2016,mehta_multiion_2020,niffenegger_multiwav_2020,kwon_multisite_optical_2023}. 

\begin{acknowledgments}
The authors are grateful for the long-standing collaboration with Sandia National Laboratories on developing  and fabricating the ion trap used in this work. We thank Adam Brandt and Ingrid Zimmermann for helpful comments on the manuscript. NKL and JFN are associates in the Professional Research Experience Program (PREP) operated jointly by NIST and the University of Colorado. This work was supported by the NIST Quantum Information Program. 

JFN performed the experiments with assistance from NKL and DL. NKL wrote the manuscript and analyzed data with assistance from JFN. All authors contributed to experimental design, apparatus maintenance, and manuscript editing. ACW, DHS, and DL secured funding for the work. DL and DHS supervised the work.
\end{acknowledgments}
\bibliography{refs}{}

\begin{thebibliography}{60}%
\makeatletter
\providecommand \@ifxundefined [1]{%
 \@ifx{#1\undefined}
}%
\providecommand \@ifnum [1]{%
 \ifnum #1\expandafter \@firstoftwo
 \else \expandafter \@secondoftwo
 \fi
}%
\providecommand \@ifx [1]{%
 \ifx #1\expandafter \@firstoftwo
 \else \expandafter \@secondoftwo
 \fi
}%
\providecommand \natexlab [1]{#1}%
\providecommand \enquote  [1]{``#1''}%
\providecommand \bibnamefont  [1]{#1}%
\providecommand \bibfnamefont [1]{#1}%
\providecommand \citenamefont [1]{#1}%
\providecommand \href@noop [0]{\@secondoftwo}%
\providecommand \href [0]{\begingroup \@sanitize@url \@href}%
\providecommand \@href[1]{\@@startlink{#1}\@@href}%
\providecommand \@@href[1]{\endgroup#1\@@endlink}%
\providecommand \@sanitize@url [0]{\catcode `\\12\catcode `\$12\catcode
  `\&12\catcode `\#12\catcode `\^12\catcode `\_12\catcode `\%12\relax}%
\providecommand \@@startlink[1]{}%
\providecommand \@@endlink[0]{}%
\providecommand \url  [0]{\begingroup\@sanitize@url \@url }%
\providecommand \@url [1]{\endgroup\@href {#1}{\urlprefix }}%
\providecommand \urlprefix  [0]{URL }%
\providecommand \Eprint [0]{\href }%
\providecommand \doibase [0]{https://doi.org/}%
\providecommand \selectlanguage [0]{\@gobble}%
\providecommand \bibinfo  [0]{\@secondoftwo}%
\providecommand \bibfield  [0]{\@secondoftwo}%
\providecommand \translation [1]{[#1]}%
\providecommand \BibitemOpen [0]{}%
\providecommand \bibitemStop [0]{}%
\providecommand \bibitemNoStop [0]{.\EOS\space}%
\providecommand \EOS [0]{\spacefactor3000\relax}%
\providecommand \BibitemShut  [1]{\csname bibitem#1\endcsname}%
\let\auto@bib@innerbib\@empty
\bibitem [{\citenamefont {Ludlow}\ \emph {et~al.}(2015)\citenamefont {Ludlow},
  \citenamefont {Boyd}, \citenamefont {Ye}, \citenamefont {Peik},\ and\
  \citenamefont {Schmidt}}]{ludlow_atomicclocks_2015}%
  \BibitemOpen
  \bibfield  {author} {\bibinfo {author} {\bibfnamefont {A.~D.}\ \bibnamefont
  {Ludlow}}, \bibinfo {author} {\bibfnamefont {M.~M.}\ \bibnamefont {Boyd}},
  \bibinfo {author} {\bibfnamefont {J.}~\bibnamefont {Ye}}, \bibinfo {author}
  {\bibfnamefont {E.}~\bibnamefont {Peik}},\ and\ \bibinfo {author}
  {\bibfnamefont {P.~O.}\ \bibnamefont {Schmidt}},\ }\href
  {https://doi.org/10.1103/RevModPhys.87.637} {\bibfield  {journal} {\bibinfo
  {journal} {Rev. Mod. Phys.}\ }\textbf {\bibinfo {volume} {87}},\ \bibinfo
  {pages} {637} (\bibinfo {year} {2015})}\BibitemShut {NoStop}%
\bibitem [{\citenamefont {Huntemann}\ \emph {et~al.}(2016)\citenamefont
  {Huntemann}, \citenamefont {Sanner}, \citenamefont {Lipphardt}, \citenamefont
  {Tamm},\ and\ \citenamefont {Peik}}]{huntemann_clock_2016}%
  \BibitemOpen
  \bibfield  {author} {\bibinfo {author} {\bibfnamefont {N.}~\bibnamefont
  {Huntemann}}, \bibinfo {author} {\bibfnamefont {C.}~\bibnamefont {Sanner}},
  \bibinfo {author} {\bibfnamefont {B.}~\bibnamefont {Lipphardt}}, \bibinfo
  {author} {\bibfnamefont {C.}~\bibnamefont {Tamm}},\ and\ \bibinfo {author}
  {\bibfnamefont {E.}~\bibnamefont {Peik}},\ }\href
  {https://doi.org/10.1103/PhysRevLett.116.063001} {\bibfield  {journal}
  {\bibinfo  {journal} {Phys. Rev. Lett.}\ }\textbf {\bibinfo {volume} {116}},\
  \bibinfo {pages} {063001} (\bibinfo {year} {2016})}\BibitemShut {NoStop}%
\bibitem [{\citenamefont {Brewer}\ \emph {et~al.}(2019)\citenamefont {Brewer},
  \citenamefont {Chen}, \citenamefont {Hankin}, \citenamefont {Clements},
  \citenamefont {Chou}, \citenamefont {Wineland}, \citenamefont {Hume},\ and\
  \citenamefont {Leibrandt}}]{brewer_prl_2019}%
  \BibitemOpen
  \bibfield  {author} {\bibinfo {author} {\bibfnamefont {S.~M.}\ \bibnamefont
  {Brewer}}, \bibinfo {author} {\bibfnamefont {J.-S.}\ \bibnamefont {Chen}},
  \bibinfo {author} {\bibfnamefont {A.~M.}\ \bibnamefont {Hankin}}, \bibinfo
  {author} {\bibfnamefont {E.~R.}\ \bibnamefont {Clements}}, \bibinfo {author}
  {\bibfnamefont {C.~W.}\ \bibnamefont {Chou}}, \bibinfo {author}
  {\bibfnamefont {D.~J.}\ \bibnamefont {Wineland}}, \bibinfo {author}
  {\bibfnamefont {D.~B.}\ \bibnamefont {Hume}},\ and\ \bibinfo {author}
  {\bibfnamefont {D.~R.}\ \bibnamefont {Leibrandt}},\ }\href
  {https://doi.org/10.1103/PhysRevLett.123.033201} {\bibfield  {journal}
  {\bibinfo  {journal} {Phys. Rev. Lett.}\ }\textbf {\bibinfo {volume} {123}},\
  \bibinfo {pages} {033201} (\bibinfo {year} {2019})}\BibitemShut {NoStop}%
\bibitem [{\citenamefont {Burt}\ \emph {et~al.}(2021)\citenamefont {Burt},
  \citenamefont {Prestage}, \citenamefont {Tjoelker}, \citenamefont {Enzer},
  \citenamefont {Kuang}, \citenamefont {Murphy}, \citenamefont {Robison},
  \citenamefont {Seubert}, \citenamefont {Wang},\ and\ \citenamefont
  {Ely}}]{burt_space_2021}%
  \BibitemOpen
  \bibfield  {author} {\bibinfo {author} {\bibfnamefont {E.~A.}\ \bibnamefont
  {Burt}}, \bibinfo {author} {\bibfnamefont {J.~D.}\ \bibnamefont {Prestage}},
  \bibinfo {author} {\bibfnamefont {R.~L.}\ \bibnamefont {Tjoelker}}, \bibinfo
  {author} {\bibfnamefont {D.~G.}\ \bibnamefont {Enzer}}, \bibinfo {author}
  {\bibfnamefont {D.}~\bibnamefont {Kuang}}, \bibinfo {author} {\bibfnamefont
  {D.~W.}\ \bibnamefont {Murphy}}, \bibinfo {author} {\bibfnamefont {D.~E.}\
  \bibnamefont {Robison}}, \bibinfo {author} {\bibfnamefont {J.~M.}\
  \bibnamefont {Seubert}}, \bibinfo {author} {\bibfnamefont {R.~T.}\
  \bibnamefont {Wang}},\ and\ \bibinfo {author} {\bibfnamefont {T.~A.}\
  \bibnamefont {Ely}},\ }\href {https://doi.org/10.1038/s41586-021-03571-7}
  {\bibfield  {journal} {\bibinfo  {journal} {Nature}\ }\textbf {\bibinfo
  {volume} {595}},\ \bibinfo {pages} {43} (\bibinfo {year} {2021})}\BibitemShut
  {NoStop}%
\bibitem [{\citenamefont {Blatt}\ and\ \citenamefont
  {Roos}(2012)}]{blatt_naturep_2012}%
  \BibitemOpen
  \bibfield  {author} {\bibinfo {author} {\bibfnamefont {R.}~\bibnamefont
  {Blatt}}\ and\ \bibinfo {author} {\bibfnamefont {C.~F.}\ \bibnamefont
  {Roos}},\ }\href {https://doi.org/10.1038/nphys2252} {\bibfield  {journal}
  {\bibinfo  {journal} {Nat. Phys.}\ }\textbf {\bibinfo {volume} {8}},\
  \bibinfo {pages} {277} (\bibinfo {year} {2012})}\BibitemShut {NoStop}%
\bibitem [{\citenamefont {Hempel}\ \emph {et~al.}(2018)\citenamefont {Hempel},
  \citenamefont {Maier}, \citenamefont {Romero}, \citenamefont {McClean},
  \citenamefont {Monz}, \citenamefont {Shen}, \citenamefont {Jurcevic},
  \citenamefont {Lanyon}, \citenamefont {Love}, \citenamefont {Babbush},
  \citenamefont {Aspuru-Guzik}, \citenamefont {Blatt},\ and\ \citenamefont
  {Roos}}]{hempel_qchem_2018}%
  \BibitemOpen
  \bibfield  {author} {\bibinfo {author} {\bibfnamefont {C.}~\bibnamefont
  {Hempel}}, \bibinfo {author} {\bibfnamefont {C.}~\bibnamefont {Maier}},
  \bibinfo {author} {\bibfnamefont {J.}~\bibnamefont {Romero}}, \bibinfo
  {author} {\bibfnamefont {J.}~\bibnamefont {McClean}}, \bibinfo {author}
  {\bibfnamefont {T.}~\bibnamefont {Monz}}, \bibinfo {author} {\bibfnamefont
  {H.}~\bibnamefont {Shen}}, \bibinfo {author} {\bibfnamefont {P.}~\bibnamefont
  {Jurcevic}}, \bibinfo {author} {\bibfnamefont {B.~P.}\ \bibnamefont
  {Lanyon}}, \bibinfo {author} {\bibfnamefont {P.}~\bibnamefont {Love}},
  \bibinfo {author} {\bibfnamefont {R.}~\bibnamefont {Babbush}}, \bibinfo
  {author} {\bibfnamefont {A.}~\bibnamefont {Aspuru-Guzik}}, \bibinfo {author}
  {\bibfnamefont {R.}~\bibnamefont {Blatt}},\ and\ \bibinfo {author}
  {\bibfnamefont {C.~F.}\ \bibnamefont {Roos}},\ }\href
  {https://doi.org/10.1103/PhysRevX.8.031022} {\bibfield  {journal} {\bibinfo
  {journal} {Phys. Rev. X}\ }\textbf {\bibinfo {volume} {8}},\ \bibinfo {pages}
  {031022} (\bibinfo {year} {2018})}\BibitemShut {NoStop}%
\bibitem [{\citenamefont {Monroe}\ \emph {et~al.}(2021)\citenamefont {Monroe},
  \citenamefont {Campbell}, \citenamefont {Duan}, \citenamefont {Gong},
  \citenamefont {Gorshkov}, \citenamefont {Hess}, \citenamefont {Islam},
  \citenamefont {Kim}, \citenamefont {Linke}, \citenamefont {Pagano},
  \citenamefont {Richerme}, \citenamefont {Senko},\ and\ \citenamefont
  {Yao}}]{monroe_rmp_2021}%
  \BibitemOpen
  \bibfield  {author} {\bibinfo {author} {\bibfnamefont {C.}~\bibnamefont
  {Monroe}}, \bibinfo {author} {\bibfnamefont {W.~C.}\ \bibnamefont
  {Campbell}}, \bibinfo {author} {\bibfnamefont {L.-M.}\ \bibnamefont {Duan}},
  \bibinfo {author} {\bibfnamefont {Z.-X.}\ \bibnamefont {Gong}}, \bibinfo
  {author} {\bibfnamefont {A.~V.}\ \bibnamefont {Gorshkov}}, \bibinfo {author}
  {\bibfnamefont {P.~W.}\ \bibnamefont {Hess}}, \bibinfo {author}
  {\bibfnamefont {R.}~\bibnamefont {Islam}}, \bibinfo {author} {\bibfnamefont
  {K.}~\bibnamefont {Kim}}, \bibinfo {author} {\bibfnamefont {N.~M.}\
  \bibnamefont {Linke}}, \bibinfo {author} {\bibfnamefont {G.}~\bibnamefont
  {Pagano}}, \bibinfo {author} {\bibfnamefont {P.}~\bibnamefont {Richerme}},
  \bibinfo {author} {\bibfnamefont {C.}~\bibnamefont {Senko}},\ and\ \bibinfo
  {author} {\bibfnamefont {N.~Y.}\ \bibnamefont {Yao}},\ }\href
  {https://doi.org/10.1103/RevModPhys.93.025001} {\bibfield  {journal}
  {\bibinfo  {journal} {Rev. Mod. Phys.}\ }\textbf {\bibinfo {volume} {93}},\
  \bibinfo {pages} {025001} (\bibinfo {year} {2021})}\BibitemShut {NoStop}%
\bibitem [{\citenamefont {Shapira}\ \emph {et~al.}(2023)\citenamefont
  {Shapira}, \citenamefont {Manovitz}, \citenamefont {Akerman}, \citenamefont
  {Stern},\ and\ \citenamefont {Ozeri}}]{shapira_timereversal_2023}%
  \BibitemOpen
  \bibfield  {author} {\bibinfo {author} {\bibfnamefont {Y.}~\bibnamefont
  {Shapira}}, \bibinfo {author} {\bibfnamefont {T.}~\bibnamefont {Manovitz}},
  \bibinfo {author} {\bibfnamefont {N.}~\bibnamefont {Akerman}}, \bibinfo
  {author} {\bibfnamefont {A.}~\bibnamefont {Stern}},\ and\ \bibinfo {author}
  {\bibfnamefont {R.}~\bibnamefont {Ozeri}},\ }\href
  {https://doi.org/10.1103/PhysRevX.13.021021} {\bibfield  {journal} {\bibinfo
  {journal} {Phys. Rev. X}\ }\textbf {\bibinfo {volume} {13}},\ \bibinfo
  {pages} {021021} (\bibinfo {year} {2023})}\BibitemShut {NoStop}%
\bibitem [{\citenamefont {Blatt}\ and\ \citenamefont
  {Wineland}(2008)}]{blatt_nature_2008}%
  \BibitemOpen
  \bibfield  {author} {\bibinfo {author} {\bibfnamefont {R.}~\bibnamefont
  {Blatt}}\ and\ \bibinfo {author} {\bibfnamefont {D.}~\bibnamefont
  {Wineland}},\ }\href@noop {} {\bibfield  {journal} {\bibinfo  {journal}
  {Nature}\ }\textbf {\bibinfo {volume} {453}},\ \bibinfo {pages} {1008}
  (\bibinfo {year} {2008})}\BibitemShut {NoStop}%
\bibitem [{\citenamefont {Bruzewicz}\ \emph {et~al.}(2019)\citenamefont
  {Bruzewicz}, \citenamefont {Chiaverini}, \citenamefont {McConnell},\ and\
  \citenamefont {Sage}}]{Bruzewicz2019}%
  \BibitemOpen
  \bibfield  {author} {\bibinfo {author} {\bibfnamefont {C.~D.}\ \bibnamefont
  {Bruzewicz}}, \bibinfo {author} {\bibfnamefont {J.}~\bibnamefont
  {Chiaverini}}, \bibinfo {author} {\bibfnamefont {R.}~\bibnamefont
  {McConnell}},\ and\ \bibinfo {author} {\bibfnamefont {J.~M.}\ \bibnamefont
  {Sage}},\ }\href {https://doi.org/10.1063/1.5088164} {\bibfield  {journal}
  {\bibinfo  {journal} {Appl. Phys. Rev.}\ }\textbf {\bibinfo {volume} {6}},\
  \bibinfo {pages} {021314} (\bibinfo {year} {2019})}\BibitemShut {NoStop}%
\bibitem [{\citenamefont {Erhard}\ \emph {et~al.}(2021)\citenamefont {Erhard},
  \citenamefont {Poulsen~Nautrup}, \citenamefont {Meth}, \citenamefont
  {Postler}, \citenamefont {Stricker}, \citenamefont {Stadler}, \citenamefont
  {Negnevitsky}, \citenamefont {Ringbauer}, \citenamefont {Schindler},
  \citenamefont {Briegel}, \citenamefont {Blatt}, \citenamefont {Friis},\ and\
  \citenamefont {Monz}}]{erhard_latsurgery_2021}%
  \BibitemOpen
  \bibfield  {author} {\bibinfo {author} {\bibfnamefont {A.}~\bibnamefont
  {Erhard}}, \bibinfo {author} {\bibfnamefont {H.}~\bibnamefont
  {Poulsen~Nautrup}}, \bibinfo {author} {\bibfnamefont {M.}~\bibnamefont
  {Meth}}, \bibinfo {author} {\bibfnamefont {L.}~\bibnamefont {Postler}},
  \bibinfo {author} {\bibfnamefont {R.}~\bibnamefont {Stricker}}, \bibinfo
  {author} {\bibfnamefont {M.}~\bibnamefont {Stadler}}, \bibinfo {author}
  {\bibfnamefont {V.}~\bibnamefont {Negnevitsky}}, \bibinfo {author}
  {\bibfnamefont {M.}~\bibnamefont {Ringbauer}}, \bibinfo {author}
  {\bibfnamefont {P.}~\bibnamefont {Schindler}}, \bibinfo {author}
  {\bibfnamefont {H.~J.}\ \bibnamefont {Briegel}}, \bibinfo {author}
  {\bibfnamefont {R.}~\bibnamefont {Blatt}}, \bibinfo {author} {\bibfnamefont
  {N.}~\bibnamefont {Friis}},\ and\ \bibinfo {author} {\bibfnamefont
  {T.}~\bibnamefont {Monz}},\ }\href
  {https://doi.org/10.1038/s41586-020-03079-6} {\bibfield  {journal} {\bibinfo
  {journal} {Nature}\ }\textbf {\bibinfo {volume} {589}},\ \bibinfo {pages}
  {220} (\bibinfo {year} {2021})}\BibitemShut {NoStop}%
\bibitem [{\citenamefont {Egan}\ \emph {et~al.}(2021)\citenamefont {Egan},
  \citenamefont {Debroy}, \citenamefont {Noel}, \citenamefont {Risinger},
  \citenamefont {Zhu}, \citenamefont {Biswas}, \citenamefont {Newman},
  \citenamefont {Li}, \citenamefont {Brown}, \citenamefont {Cetina},\ and\
  \citenamefont {Monroe}}]{egan_ftc_2021}%
  \BibitemOpen
  \bibfield  {author} {\bibinfo {author} {\bibfnamefont {L.}~\bibnamefont
  {Egan}}, \bibinfo {author} {\bibfnamefont {D.~M.}\ \bibnamefont {Debroy}},
  \bibinfo {author} {\bibfnamefont {C.}~\bibnamefont {Noel}}, \bibinfo {author}
  {\bibfnamefont {A.}~\bibnamefont {Risinger}}, \bibinfo {author}
  {\bibfnamefont {D.}~\bibnamefont {Zhu}}, \bibinfo {author} {\bibfnamefont
  {D.}~\bibnamefont {Biswas}}, \bibinfo {author} {\bibfnamefont
  {M.}~\bibnamefont {Newman}}, \bibinfo {author} {\bibfnamefont
  {M.}~\bibnamefont {Li}}, \bibinfo {author} {\bibfnamefont {K.~R.}\
  \bibnamefont {Brown}}, \bibinfo {author} {\bibfnamefont {M.}~\bibnamefont
  {Cetina}},\ and\ \bibinfo {author} {\bibfnamefont {C.}~\bibnamefont
  {Monroe}},\ }\href {https://doi.org/10.1038/s41586-021-03928-y} {\bibfield
  {journal} {\bibinfo  {journal} {Nature}\ }\textbf {\bibinfo {volume} {598}},\
  \bibinfo {pages} {281} (\bibinfo {year} {2021})}\BibitemShut {NoStop}%
\bibitem [{\citenamefont {Ryan-Anderson}\ \emph {et~al.}(2022)\citenamefont
  {Ryan-Anderson}, \citenamefont {Brown}, \citenamefont {Allman}, \citenamefont
  {Arkin}, \citenamefont {Asa-Attuah}, \citenamefont {Baldwin}, \citenamefont
  {Berg}, \citenamefont {Bohnet}, \citenamefont {Braxton}, \citenamefont
  {Burdick}, \citenamefont {Campora}, \citenamefont {Chernoguzov},
  \citenamefont {Esposito}, \citenamefont {Evans}, \citenamefont {Francois}
  \emph {et~al.}}]{ryananderson_qec_2022}%
  \BibitemOpen
  \bibfield  {author} {\bibinfo {author} {\bibfnamefont {C.}~\bibnamefont
  {Ryan-Anderson}}, \bibinfo {author} {\bibfnamefont {N.~C.}\ \bibnamefont
  {Brown}}, \bibinfo {author} {\bibfnamefont {M.~S.}\ \bibnamefont {Allman}},
  \bibinfo {author} {\bibfnamefont {B.}~\bibnamefont {Arkin}}, \bibinfo
  {author} {\bibfnamefont {G.}~\bibnamefont {Asa-Attuah}}, \bibinfo {author}
  {\bibfnamefont {C.}~\bibnamefont {Baldwin}}, \bibinfo {author} {\bibfnamefont
  {J.}~\bibnamefont {Berg}}, \bibinfo {author} {\bibfnamefont {J.~G.}\
  \bibnamefont {Bohnet}}, \bibinfo {author} {\bibfnamefont {S.}~\bibnamefont
  {Braxton}}, \bibinfo {author} {\bibfnamefont {N.}~\bibnamefont {Burdick}},
  \bibinfo {author} {\bibfnamefont {J.~P.}\ \bibnamefont {Campora}}, \bibinfo
  {author} {\bibfnamefont {A.}~\bibnamefont {Chernoguzov}}, \bibinfo {author}
  {\bibfnamefont {J.}~\bibnamefont {Esposito}}, \bibinfo {author}
  {\bibfnamefont {B.}~\bibnamefont {Evans}}, \bibinfo {author} {\bibfnamefont
  {D.}~\bibnamefont {Francois}}, \emph {et~al.},\ }\Eprint
  {https://arxiv.org/abs/2208.01863} {arXiv:2208.01863 [quant-ph]}  (\bibinfo
  {year} {2022})\BibitemShut {NoStop}%
\bibitem [{\citenamefont {Postler}\ \emph {et~al.}(2022)\citenamefont
  {Postler}, \citenamefont {Heu$\beta$en}, \citenamefont {Pogorelov},
  \citenamefont {Rispler}, \citenamefont {Feldker}, \citenamefont {Meth},
  \citenamefont {Marciniak}, \citenamefont {Stricker}, \citenamefont
  {Ringbauer}, \citenamefont {Blatt}, \citenamefont {Schindler}, \citenamefont
  {M{\"u}ller},\ and\ \citenamefont {Monz}}]{postler_ftuqgate_2022}%
  \BibitemOpen
  \bibfield  {author} {\bibinfo {author} {\bibfnamefont {L.}~\bibnamefont
  {Postler}}, \bibinfo {author} {\bibfnamefont {S.}~\bibnamefont
  {Heu$\beta$en}}, \bibinfo {author} {\bibfnamefont {I.}~\bibnamefont
  {Pogorelov}}, \bibinfo {author} {\bibfnamefont {M.}~\bibnamefont {Rispler}},
  \bibinfo {author} {\bibfnamefont {T.}~\bibnamefont {Feldker}}, \bibinfo
  {author} {\bibfnamefont {M.}~\bibnamefont {Meth}}, \bibinfo {author}
  {\bibfnamefont {C.~D.}\ \bibnamefont {Marciniak}}, \bibinfo {author}
  {\bibfnamefont {R.}~\bibnamefont {Stricker}}, \bibinfo {author}
  {\bibfnamefont {M.}~\bibnamefont {Ringbauer}}, \bibinfo {author}
  {\bibfnamefont {R.}~\bibnamefont {Blatt}}, \bibinfo {author} {\bibfnamefont
  {P.}~\bibnamefont {Schindler}}, \bibinfo {author} {\bibfnamefont
  {M.}~\bibnamefont {M{\"u}ller}},\ and\ \bibinfo {author} {\bibfnamefont
  {T.}~\bibnamefont {Monz}},\ }\href
  {https://doi.org/10.1038/s41586-022-04721-1} {\bibfield  {journal} {\bibinfo
  {journal} {Nature}\ }\textbf {\bibinfo {volume} {605}},\ \bibinfo {pages}
  {675} (\bibinfo {year} {2022})}\BibitemShut {NoStop}%
\bibitem [{\citenamefont {Godun}\ \emph {et~al.}(2014)\citenamefont {Godun},
  \citenamefont {Nisbet-Jones}, \citenamefont {Jones}, \citenamefont {King},
  \citenamefont {Johnson}, \citenamefont {Margolis}, \citenamefont {Szymaniec},
  \citenamefont {Lea}, \citenamefont {Bongs},\ and\ \citenamefont
  {Gill}}]{godun_timevar_2014}%
  \BibitemOpen
  \bibfield  {author} {\bibinfo {author} {\bibfnamefont {R.~M.}\ \bibnamefont
  {Godun}}, \bibinfo {author} {\bibfnamefont {P.~B.~R.}\ \bibnamefont
  {Nisbet-Jones}}, \bibinfo {author} {\bibfnamefont {J.~M.}\ \bibnamefont
  {Jones}}, \bibinfo {author} {\bibfnamefont {S.~A.}\ \bibnamefont {King}},
  \bibinfo {author} {\bibfnamefont {L.~A.~M.}\ \bibnamefont {Johnson}},
  \bibinfo {author} {\bibfnamefont {H.~S.}\ \bibnamefont {Margolis}}, \bibinfo
  {author} {\bibfnamefont {K.}~\bibnamefont {Szymaniec}}, \bibinfo {author}
  {\bibfnamefont {S.~N.}\ \bibnamefont {Lea}}, \bibinfo {author} {\bibfnamefont
  {K.}~\bibnamefont {Bongs}},\ and\ \bibinfo {author} {\bibfnamefont
  {P.}~\bibnamefont {Gill}},\ }\href
  {https://doi.org/10.1103/PhysRevLett.113.210801} {\bibfield  {journal}
  {\bibinfo  {journal} {Phys. Rev. Lett.}\ }\textbf {\bibinfo {volume} {113}},\
  \bibinfo {pages} {210801} (\bibinfo {year} {2014})}\BibitemShut {NoStop}%
\bibitem [{\citenamefont {Huntemann}\ \emph {et~al.}(2014)\citenamefont
  {Huntemann}, \citenamefont {Lipphardt}, \citenamefont {Tamm}, \citenamefont
  {Gerginov}, \citenamefont {Weyers},\ and\ \citenamefont
  {Peik}}]{huntemann_massvar_2014}%
  \BibitemOpen
  \bibfield  {author} {\bibinfo {author} {\bibfnamefont {N.}~\bibnamefont
  {Huntemann}}, \bibinfo {author} {\bibfnamefont {B.}~\bibnamefont
  {Lipphardt}}, \bibinfo {author} {\bibfnamefont {C.}~\bibnamefont {Tamm}},
  \bibinfo {author} {\bibfnamefont {V.}~\bibnamefont {Gerginov}}, \bibinfo
  {author} {\bibfnamefont {S.}~\bibnamefont {Weyers}},\ and\ \bibinfo {author}
  {\bibfnamefont {E.}~\bibnamefont {Peik}},\ }\href
  {https://doi.org/10.1103/PhysRevLett.113.210802} {\bibfield  {journal}
  {\bibinfo  {journal} {Phys. Rev. Lett.}\ }\textbf {\bibinfo {volume} {113}},\
  \bibinfo {pages} {210802} (\bibinfo {year} {2014})}\BibitemShut {NoStop}%
\bibitem [{\citenamefont {Pruttivarasin}\ \emph {et~al.}(2015)\citenamefont
  {Pruttivarasin}, \citenamefont {Ramm}, \citenamefont {Porsev}, \citenamefont
  {Tupitsyn}, \citenamefont {Safronova}, \citenamefont {Hohensee},\ and\
  \citenamefont {H{\"a}ffner}}]{pruttivarasin_nature_2015}%
  \BibitemOpen
  \bibfield  {author} {\bibinfo {author} {\bibfnamefont {T.}~\bibnamefont
  {Pruttivarasin}}, \bibinfo {author} {\bibfnamefont {M.}~\bibnamefont {Ramm}},
  \bibinfo {author} {\bibfnamefont {S.~G.}\ \bibnamefont {Porsev}}, \bibinfo
  {author} {\bibfnamefont {I.~I.}\ \bibnamefont {Tupitsyn}}, \bibinfo {author}
  {\bibfnamefont {M.~S.}\ \bibnamefont {Safronova}}, \bibinfo {author}
  {\bibfnamefont {M.~A.}\ \bibnamefont {Hohensee}},\ and\ \bibinfo {author}
  {\bibfnamefont {H.}~\bibnamefont {H{\"a}ffner}},\ }\href@noop {} {\bibfield
  {journal} {\bibinfo  {journal} {Nature}\ }\textbf {\bibinfo {volume} {517}},\
  \bibinfo {pages} {592} (\bibinfo {year} {2015})}\BibitemShut {NoStop}%
\bibitem [{\citenamefont {Dreissen}\ \emph {et~al.}(2022)\citenamefont
  {Dreissen}, \citenamefont {Yeh}, \citenamefont {F{\"u}rst}, \citenamefont
  {Grensemann},\ and\ \citenamefont
  {Mehlst{\"a}ubler}}]{dreissen_lorentz_2022}%
  \BibitemOpen
  \bibfield  {author} {\bibinfo {author} {\bibfnamefont {L.~S.}\ \bibnamefont
  {Dreissen}}, \bibinfo {author} {\bibfnamefont {C.-H.}\ \bibnamefont {Yeh}},
  \bibinfo {author} {\bibfnamefont {H.~A.}\ \bibnamefont {F{\"u}rst}}, \bibinfo
  {author} {\bibfnamefont {K.~C.}\ \bibnamefont {Grensemann}},\ and\ \bibinfo
  {author} {\bibfnamefont {T.~E.}\ \bibnamefont {Mehlst{\"a}ubler}},\ }\href
  {https://doi.org/10.1038/s41467-022-34818-0} {\bibfield  {journal} {\bibinfo
  {journal} {Nat. Comm.}\ }\textbf {\bibinfo {volume} {13}},\ \bibinfo {pages}
  {7314} (\bibinfo {year} {2022})}\BibitemShut {NoStop}%
\bibitem [{\citenamefont {Kozlov}\ \emph {et~al.}(2018)\citenamefont {Kozlov},
  \citenamefont {Safronova}, \citenamefont {Crespo L\'opez-Urrutia},\ and\
  \citenamefont {Schmidt}}]{kozlov_hci_2018}%
  \BibitemOpen
  \bibfield  {author} {\bibinfo {author} {\bibfnamefont {M.~G.}\ \bibnamefont
  {Kozlov}}, \bibinfo {author} {\bibfnamefont {M.~S.}\ \bibnamefont
  {Safronova}}, \bibinfo {author} {\bibfnamefont {J.~R.}\ \bibnamefont {Crespo
  L\'opez-Urrutia}},\ and\ \bibinfo {author} {\bibfnamefont {P.~O.}\
  \bibnamefont {Schmidt}},\ }\href
  {https://doi.org/10.1103/RevModPhys.90.045005} {\bibfield  {journal}
  {\bibinfo  {journal} {Rev. Mod. Phys.}\ }\textbf {\bibinfo {volume} {90}},\
  \bibinfo {pages} {045005} (\bibinfo {year} {2018})}\BibitemShut {NoStop}%
\bibitem [{\citenamefont {Arrowsmith-Kron}\ \emph {et~al.}(2023)\citenamefont
  {Arrowsmith-Kron}, \citenamefont {Athanasakis-Kaklamanakis}, \citenamefont
  {Au}, \citenamefont {Ballof}, \citenamefont {Berger}, \citenamefont
  {Borschevsky}, \citenamefont {Breier}, \citenamefont {Buchinger},
  \citenamefont {Budker}, \citenamefont {Caldwell}, \citenamefont {Charles},
  \citenamefont {Dattani}, \citenamefont {de~Groote}, \citenamefont {DeMille},
  \citenamefont {Dickel} \emph {et~al.}}]{arrowsmith_radmol_2023}%
  \BibitemOpen
  \bibfield  {author} {\bibinfo {author} {\bibfnamefont {G.}~\bibnamefont
  {Arrowsmith-Kron}}, \bibinfo {author} {\bibfnamefont {M.}~\bibnamefont
  {Athanasakis-Kaklamanakis}}, \bibinfo {author} {\bibfnamefont
  {M.}~\bibnamefont {Au}}, \bibinfo {author} {\bibfnamefont {J.}~\bibnamefont
  {Ballof}}, \bibinfo {author} {\bibfnamefont {R.}~\bibnamefont {Berger}},
  \bibinfo {author} {\bibfnamefont {A.}~\bibnamefont {Borschevsky}}, \bibinfo
  {author} {\bibfnamefont {A.~A.}\ \bibnamefont {Breier}}, \bibinfo {author}
  {\bibfnamefont {F.}~\bibnamefont {Buchinger}}, \bibinfo {author}
  {\bibfnamefont {D.}~\bibnamefont {Budker}}, \bibinfo {author} {\bibfnamefont
  {L.}~\bibnamefont {Caldwell}}, \bibinfo {author} {\bibfnamefont
  {C.}~\bibnamefont {Charles}}, \bibinfo {author} {\bibfnamefont
  {N.}~\bibnamefont {Dattani}}, \bibinfo {author} {\bibfnamefont {R.~P.}\
  \bibnamefont {de~Groote}}, \bibinfo {author} {\bibfnamefont {D.}~\bibnamefont
  {DeMille}}, \bibinfo {author} {\bibfnamefont {T.}~\bibnamefont {Dickel}},
  \emph {et~al.},\ }\Eprint {https://arxiv.org/abs/2302.02165}
  {arXiv:2302.02165 [nucl-ex]}  (\bibinfo {year} {2023})\BibitemShut {NoStop}%
\bibitem [{\citenamefont {Wang}\ \emph {et~al.}(2020)\citenamefont {Wang},
  \citenamefont {Crain}, \citenamefont {Fang}, \citenamefont {Zhang},
  \citenamefont {Huang}, \citenamefont {Liang}, \citenamefont {Leung},
  \citenamefont {Brown},\ and\ \citenamefont {Kim}}]{wang_twoqubit_MEMS_2020}%
  \BibitemOpen
  \bibfield  {author} {\bibinfo {author} {\bibfnamefont {Y.}~\bibnamefont
  {Wang}}, \bibinfo {author} {\bibfnamefont {S.}~\bibnamefont {Crain}},
  \bibinfo {author} {\bibfnamefont {C.}~\bibnamefont {Fang}}, \bibinfo {author}
  {\bibfnamefont {B.}~\bibnamefont {Zhang}}, \bibinfo {author} {\bibfnamefont
  {S.}~\bibnamefont {Huang}}, \bibinfo {author} {\bibfnamefont
  {Q.}~\bibnamefont {Liang}}, \bibinfo {author} {\bibfnamefont {P.~H.}\
  \bibnamefont {Leung}}, \bibinfo {author} {\bibfnamefont {K.~R.}\ \bibnamefont
  {Brown}},\ and\ \bibinfo {author} {\bibfnamefont {J.}~\bibnamefont {Kim}},\
  }\href {https://doi.org/10.1103/PhysRevLett.125.150505} {\bibfield  {journal}
  {\bibinfo  {journal} {Phys. Rev. Lett.}\ }\textbf {\bibinfo {volume} {125}},\
  \bibinfo {pages} {150505} (\bibinfo {year} {2020})}\BibitemShut {NoStop}%
\bibitem [{\citenamefont {Kranzl}\ \emph {et~al.}(2022)\citenamefont {Kranzl},
  \citenamefont {Joshi}, \citenamefont {Maier}, \citenamefont {Brydges},
  \citenamefont {Franke}, \citenamefont {Blatt},\ and\ \citenamefont
  {Roos}}]{kranzel_strings_2022}%
  \BibitemOpen
  \bibfield  {author} {\bibinfo {author} {\bibfnamefont {F.}~\bibnamefont
  {Kranzl}}, \bibinfo {author} {\bibfnamefont {M.~K.}\ \bibnamefont {Joshi}},
  \bibinfo {author} {\bibfnamefont {C.}~\bibnamefont {Maier}}, \bibinfo
  {author} {\bibfnamefont {T.}~\bibnamefont {Brydges}}, \bibinfo {author}
  {\bibfnamefont {J.}~\bibnamefont {Franke}}, \bibinfo {author} {\bibfnamefont
  {R.}~\bibnamefont {Blatt}},\ and\ \bibinfo {author} {\bibfnamefont {C.~F.}\
  \bibnamefont {Roos}},\ }\href {https://doi.org/10.1103/PhysRevA.105.052426}
  {\bibfield  {journal} {\bibinfo  {journal} {Phys. Rev. A}\ }\textbf {\bibinfo
  {volume} {105}},\ \bibinfo {pages} {052426} (\bibinfo {year}
  {2022})}\BibitemShut {NoStop}%
\bibitem [{\citenamefont {Joshi}\ \emph {et~al.}(2022)\citenamefont {Joshi},
  \citenamefont {Kranzl}, \citenamefont {Schuckert}, \citenamefont {Lovas},
  \citenamefont {Maier}, \citenamefont {Blatt}, \citenamefont {Knap},\ and\
  \citenamefont {Roos}}]{joshi_hydrodynamics_2022}%
  \BibitemOpen
  \bibfield  {author} {\bibinfo {author} {\bibfnamefont {M.~K.}\ \bibnamefont
  {Joshi}}, \bibinfo {author} {\bibfnamefont {F.}~\bibnamefont {Kranzl}},
  \bibinfo {author} {\bibfnamefont {A.}~\bibnamefont {Schuckert}}, \bibinfo
  {author} {\bibfnamefont {I.}~\bibnamefont {Lovas}}, \bibinfo {author}
  {\bibfnamefont {C.}~\bibnamefont {Maier}}, \bibinfo {author} {\bibfnamefont
  {R.}~\bibnamefont {Blatt}}, \bibinfo {author} {\bibfnamefont
  {M.}~\bibnamefont {Knap}},\ and\ \bibinfo {author} {\bibfnamefont {C.~F.}\
  \bibnamefont {Roos}},\ }\href {https://doi.org/10.1126/science.abk2400}
  {\bibfield  {journal} {\bibinfo  {journal} {Science}\ }\textbf {\bibinfo
  {volume} {376}},\ \bibinfo {pages} {720} (\bibinfo {year}
  {2022})}\BibitemShut {NoStop}%
\bibitem [{\citenamefont {Wineland}\ \emph {et~al.}(1998)\citenamefont
  {Wineland}, \citenamefont {Monroe}, \citenamefont {Itano}, \citenamefont
  {Leibfried}, \citenamefont {King},\ and\ \citenamefont
  {Meekhof}}]{wineland_experimental_1998}%
  \BibitemOpen
  \bibfield  {author} {\bibinfo {author} {\bibfnamefont {D.}~\bibnamefont
  {Wineland}}, \bibinfo {author} {\bibfnamefont {C.}~\bibnamefont {Monroe}},
  \bibinfo {author} {\bibfnamefont {W.}~\bibnamefont {Itano}}, \bibinfo
  {author} {\bibfnamefont {D.}~\bibnamefont {Leibfried}}, \bibinfo {author}
  {\bibfnamefont {B.}~\bibnamefont {King}},\ and\ \bibinfo {author}
  {\bibfnamefont {D.}~\bibnamefont {Meekhof}},\ }\href
  {https://doi.org/10.6028/jres.103.019} {\bibfield  {journal} {\bibinfo
  {journal} {J. Res. Natl. Inst. Stan.}\ }\textbf {\bibinfo {volume} {103}},\
  \bibinfo {pages} {259} (\bibinfo {year} {1998})}\BibitemShut {NoStop}%
\bibitem [{\citenamefont {Kielpinski}\ \emph {et~al.}(2002)\citenamefont
  {Kielpinski}, \citenamefont {Monroe},\ and\ \citenamefont
  {Wineland}}]{kielpinski_large_scale_2002}%
  \BibitemOpen
  \bibfield  {author} {\bibinfo {author} {\bibfnamefont {D.}~\bibnamefont
  {Kielpinski}}, \bibinfo {author} {\bibfnamefont {C.~R.}\ \bibnamefont
  {Monroe}},\ and\ \bibinfo {author} {\bibfnamefont {D.~J.}\ \bibnamefont
  {Wineland}},\ }\href {https://api.semanticscholar.org/CorpusID:4347109}
  {\bibfield  {journal} {\bibinfo  {journal} {Nature}\ }\textbf {\bibinfo
  {volume} {417}},\ \bibinfo {pages} {709} (\bibinfo {year}
  {2002})}\BibitemShut {NoStop}%
\bibitem [{\citenamefont {Wan}\ \emph {et~al.}(2019)\citenamefont {Wan},
  \citenamefont {Kienzler}, \citenamefont {Erickson}, \citenamefont {Mayer},
  \citenamefont {Tan}, \citenamefont {Wu}, \citenamefont {Vasconcelos},
  \citenamefont {Glancy}, \citenamefont {Knill}, \citenamefont {Wineland},
  \citenamefont {Wilson},\ and\ \citenamefont
  {Leibfried}}]{wan_teleport_zones_2019}%
  \BibitemOpen
  \bibfield  {author} {\bibinfo {author} {\bibfnamefont {Y.}~\bibnamefont
  {Wan}}, \bibinfo {author} {\bibfnamefont {D.}~\bibnamefont {Kienzler}},
  \bibinfo {author} {\bibfnamefont {S.~D.}\ \bibnamefont {Erickson}}, \bibinfo
  {author} {\bibfnamefont {K.~H.}\ \bibnamefont {Mayer}}, \bibinfo {author}
  {\bibfnamefont {T.~R.}\ \bibnamefont {Tan}}, \bibinfo {author} {\bibfnamefont
  {J.~J.}\ \bibnamefont {Wu}}, \bibinfo {author} {\bibfnamefont {H.~M.}\
  \bibnamefont {Vasconcelos}}, \bibinfo {author} {\bibfnamefont
  {S.}~\bibnamefont {Glancy}}, \bibinfo {author} {\bibfnamefont
  {E.}~\bibnamefont {Knill}}, \bibinfo {author} {\bibfnamefont {D.~J.}\
  \bibnamefont {Wineland}}, \bibinfo {author} {\bibfnamefont {A.~C.}\
  \bibnamefont {Wilson}},\ and\ \bibinfo {author} {\bibfnamefont
  {D.}~\bibnamefont {Leibfried}},\ }\href
  {https://doi.org/10.1126/science.aaw9415} {\bibfield  {journal} {\bibinfo
  {journal} {Science}\ }\textbf {\bibinfo {volume} {364}},\ \bibinfo {pages}
  {875} (\bibinfo {year} {2019})}\BibitemShut {NoStop}%
\bibitem [{\citenamefont {{Pino}}\ \emph {et~al.}(2021)\citenamefont {{Pino}},
  \citenamefont {{Dreiling}}, \citenamefont {{Figgatt}}, \citenamefont
  {{Gaebler}}, \citenamefont {{Moses}}, \citenamefont {{Allman}}, \citenamefont
  {{Baldwin}}, \citenamefont {{Foss-Feig}}, \citenamefont {{Hayes}},
  \citenamefont {{Mayer}}, \citenamefont {{Ryan-Anderson}},\ and\ \citenamefont
  {{Neyenhuis}}}]{pino_ccd_demonstration_2021}%
  \BibitemOpen
  \bibfield  {author} {\bibinfo {author} {\bibfnamefont {J.~M.}\ \bibnamefont
  {{Pino}}}, \bibinfo {author} {\bibfnamefont {J.~M.}\ \bibnamefont
  {{Dreiling}}}, \bibinfo {author} {\bibfnamefont {C.}~\bibnamefont
  {{Figgatt}}}, \bibinfo {author} {\bibfnamefont {J.~P.}\ \bibnamefont
  {{Gaebler}}}, \bibinfo {author} {\bibfnamefont {S.~A.}\ \bibnamefont
  {{Moses}}}, \bibinfo {author} {\bibfnamefont {M.~S.}\ \bibnamefont
  {{Allman}}}, \bibinfo {author} {\bibfnamefont {C.~H.}\ \bibnamefont
  {{Baldwin}}}, \bibinfo {author} {\bibfnamefont {M.}~\bibnamefont
  {{Foss-Feig}}}, \bibinfo {author} {\bibfnamefont {D.}~\bibnamefont
  {{Hayes}}}, \bibinfo {author} {\bibfnamefont {K.}~\bibnamefont {{Mayer}}},
  \bibinfo {author} {\bibfnamefont {C.}~\bibnamefont {{Ryan-Anderson}}},\ and\
  \bibinfo {author} {\bibfnamefont {B.}~\bibnamefont {{Neyenhuis}}},\ }\href
  {https://doi.org/10.1038/s41586-021-03318-4} {\bibfield  {journal} {\bibinfo
  {journal} {Nature}\ }\textbf {\bibinfo {volume} {592}},\ \bibinfo {pages}
  {209} (\bibinfo {year} {2021})}\BibitemShut {NoStop}%
\bibitem [{\citenamefont {Moses}\ \emph {et~al.}(2023)\citenamefont {Moses},
  \citenamefont {Baldwin}, \citenamefont {Allman}, \citenamefont {Ancona},
  \citenamefont {Ascarrunz}, \citenamefont {Barnes}, \citenamefont
  {Bartolotta}, \citenamefont {Bjork}, \citenamefont {Blanchard}, \citenamefont
  {Bohn}, \citenamefont {Bohnet}, \citenamefont {Brown}, \citenamefont
  {Burdick}, \citenamefont {Burton}, \citenamefont {Campbell} \emph
  {et~al.}}]{moses_race_2023}%
  \BibitemOpen
  \bibfield  {author} {\bibinfo {author} {\bibfnamefont {S.~A.}\ \bibnamefont
  {Moses}}, \bibinfo {author} {\bibfnamefont {C.~H.}\ \bibnamefont {Baldwin}},
  \bibinfo {author} {\bibfnamefont {M.~S.}\ \bibnamefont {Allman}}, \bibinfo
  {author} {\bibfnamefont {R.}~\bibnamefont {Ancona}}, \bibinfo {author}
  {\bibfnamefont {L.}~\bibnamefont {Ascarrunz}}, \bibinfo {author}
  {\bibfnamefont {C.}~\bibnamefont {Barnes}}, \bibinfo {author} {\bibfnamefont
  {J.}~\bibnamefont {Bartolotta}}, \bibinfo {author} {\bibfnamefont
  {B.}~\bibnamefont {Bjork}}, \bibinfo {author} {\bibfnamefont
  {P.}~\bibnamefont {Blanchard}}, \bibinfo {author} {\bibfnamefont
  {M.}~\bibnamefont {Bohn}}, \bibinfo {author} {\bibfnamefont {J.~G.}\
  \bibnamefont {Bohnet}}, \bibinfo {author} {\bibfnamefont {N.~C.}\
  \bibnamefont {Brown}}, \bibinfo {author} {\bibfnamefont {N.~Q.}\ \bibnamefont
  {Burdick}}, \bibinfo {author} {\bibfnamefont {W.~C.}\ \bibnamefont {Burton}},
  \bibinfo {author} {\bibfnamefont {S.~L.}\ \bibnamefont {Campbell}}, \emph
  {et~al.},\ }\href {https://doi.org/10.1103/PhysRevX.13.041052} {\bibfield
  {journal} {\bibinfo  {journal} {Phys. Rev. X}\ }\textbf {\bibinfo {volume}
  {13}},\ \bibinfo {pages} {041052} (\bibinfo {year} {2023})}\BibitemShut
  {NoStop}%
\bibitem [{\citenamefont {Kiesenhofer}\ \emph {et~al.}(2023)\citenamefont
  {Kiesenhofer}, \citenamefont {Hainzer}, \citenamefont {Zhdanov},
  \citenamefont {Holz}, \citenamefont {Bock}, \citenamefont {Ollikainen},\ and\
  \citenamefont {Roos}}]{kiesenhofer_2dions_2023}%
  \BibitemOpen
  \bibfield  {author} {\bibinfo {author} {\bibfnamefont {D.}~\bibnamefont
  {Kiesenhofer}}, \bibinfo {author} {\bibfnamefont {H.}~\bibnamefont
  {Hainzer}}, \bibinfo {author} {\bibfnamefont {A.}~\bibnamefont {Zhdanov}},
  \bibinfo {author} {\bibfnamefont {P.~C.}\ \bibnamefont {Holz}}, \bibinfo
  {author} {\bibfnamefont {M.}~\bibnamefont {Bock}}, \bibinfo {author}
  {\bibfnamefont {T.}~\bibnamefont {Ollikainen}},\ and\ \bibinfo {author}
  {\bibfnamefont {C.~F.}\ \bibnamefont {Roos}},\ }\href
  {https://doi.org/10.1103/PRXQuantum.4.020317} {\bibfield  {journal} {\bibinfo
   {journal} {PRX Quantum}\ }\textbf {\bibinfo {volume} {4}},\ \bibinfo {pages}
  {020317} (\bibinfo {year} {2023})}\BibitemShut {NoStop}%
\bibitem [{\citenamefont {Turchette}\ \emph {et~al.}(1998)\citenamefont
  {Turchette}, \citenamefont {Wood}, \citenamefont {King}, \citenamefont
  {Myatt}, \citenamefont {Leibfried}, \citenamefont {Itano}, \citenamefont
  {Monroe},\ and\ \citenamefont {Wineland}}]{turchette_ion_entanglement_1998}%
  \BibitemOpen
  \bibfield  {author} {\bibinfo {author} {\bibfnamefont {Q.~A.}\ \bibnamefont
  {Turchette}}, \bibinfo {author} {\bibfnamefont {C.~S.}\ \bibnamefont {Wood}},
  \bibinfo {author} {\bibfnamefont {B.~E.}\ \bibnamefont {King}}, \bibinfo
  {author} {\bibfnamefont {C.~J.}\ \bibnamefont {Myatt}}, \bibinfo {author}
  {\bibfnamefont {D.}~\bibnamefont {Leibfried}}, \bibinfo {author}
  {\bibfnamefont {W.~M.}\ \bibnamefont {Itano}}, \bibinfo {author}
  {\bibfnamefont {C.}~\bibnamefont {Monroe}},\ and\ \bibinfo {author}
  {\bibfnamefont {D.~J.}\ \bibnamefont {Wineland}},\ }\href
  {https://doi.org/10.1103/PhysRevLett.81.3631} {\bibfield  {journal} {\bibinfo
   {journal} {Phys. Rev. Lett.}\ }\textbf {\bibinfo {volume} {81}},\ \bibinfo
  {pages} {3631} (\bibinfo {year} {1998})}\BibitemShut {NoStop}%
\bibitem [{\citenamefont {Leibfried}(1999)}]{leibfried_individual_1999}%
  \BibitemOpen
  \bibfield  {author} {\bibinfo {author} {\bibfnamefont {D.}~\bibnamefont
  {Leibfried}},\ }\href {https://doi.org/10.1103/PhysRevA.60.R3335} {\bibfield
  {journal} {\bibinfo  {journal} {Phys. Rev. A}\ }\textbf {\bibinfo {volume}
  {60}},\ \bibinfo {pages} {R3335} (\bibinfo {year} {1999})}\BibitemShut
  {NoStop}%
\bibitem [{\citenamefont {Mintert}\ and\ \citenamefont
  {Wunderlich}(2001)}]{mintert_long_wave_logic_2001}%
  \BibitemOpen
  \bibfield  {author} {\bibinfo {author} {\bibfnamefont {F.}~\bibnamefont
  {Mintert}}\ and\ \bibinfo {author} {\bibfnamefont {C.}~\bibnamefont
  {Wunderlich}},\ }\href {https://doi.org/10.1103/PhysRevLett.87.257904}
  {\bibfield  {journal} {\bibinfo  {journal} {Phys. Rev. Lett.}\ }\textbf
  {\bibinfo {volume} {87}},\ \bibinfo {pages} {257904} (\bibinfo {year}
  {2001})}\BibitemShut {NoStop}%
\bibitem [{\citenamefont {Staanum}\ and\ \citenamefont
  {Drewsen}(2002)}]{staanum2002}%
  \BibitemOpen
  \bibfield  {author} {\bibinfo {author} {\bibfnamefont {P.}~\bibnamefont
  {Staanum}}\ and\ \bibinfo {author} {\bibfnamefont {M.}~\bibnamefont
  {Drewsen}},\ }\href {https://doi.org/10.1103/PhysRevA.66.040302} {\bibfield
  {journal} {\bibinfo  {journal} {Phys. Rev. A}\ }\textbf {\bibinfo {volume}
  {66}},\ \bibinfo {pages} {040302} (\bibinfo {year} {2002})}\BibitemShut
  {NoStop}%
\bibitem [{\citenamefont {Chiaverini}\ \emph {et~al.}(2004)\citenamefont
  {Chiaverini}, \citenamefont {Leibfried}, \citenamefont {Schaetz},
  \citenamefont {Barrett}, \citenamefont {Blakestad}, \citenamefont {Britton},
  \citenamefont {Itano}, \citenamefont {Jost}, \citenamefont {Knill},
  \citenamefont {Langer}, \citenamefont {Ozeri},\ and\ \citenamefont
  {Wineland}}]{chiaverini_qecc_2004}%
  \BibitemOpen
  \bibfield  {author} {\bibinfo {author} {\bibfnamefont {J.}~\bibnamefont
  {Chiaverini}}, \bibinfo {author} {\bibfnamefont {D.}~\bibnamefont
  {Leibfried}}, \bibinfo {author} {\bibfnamefont {T.}~\bibnamefont {Schaetz}},
  \bibinfo {author} {\bibfnamefont {M.~D.}\ \bibnamefont {Barrett}}, \bibinfo
  {author} {\bibfnamefont {R.~B.}\ \bibnamefont {Blakestad}}, \bibinfo {author}
  {\bibfnamefont {J.}~\bibnamefont {Britton}}, \bibinfo {author} {\bibfnamefont
  {W.~M.}\ \bibnamefont {Itano}}, \bibinfo {author} {\bibfnamefont {J.~D.}\
  \bibnamefont {Jost}}, \bibinfo {author} {\bibfnamefont {E.}~\bibnamefont
  {Knill}}, \bibinfo {author} {\bibfnamefont {C.}~\bibnamefont {Langer}},
  \bibinfo {author} {\bibfnamefont {R.}~\bibnamefont {Ozeri}},\ and\ \bibinfo
  {author} {\bibfnamefont {D.~J.}\ \bibnamefont {Wineland}},\ }\href
  {https://doi.org/10.1038/nature03074} {\bibfield  {journal} {\bibinfo
  {journal} {Nature}\ }\textbf {\bibinfo {volume} {432}},\ \bibinfo {pages}
  {602} (\bibinfo {year} {2004})}\BibitemShut {NoStop}%
\bibitem [{\citenamefont {Mc~Hugh}\ and\ \citenamefont
  {Twamley}(2005)}]{mchugh2005}%
  \BibitemOpen
  \bibfield  {author} {\bibinfo {author} {\bibfnamefont {D.}~\bibnamefont
  {Mc~Hugh}}\ and\ \bibinfo {author} {\bibfnamefont {J.}~\bibnamefont
  {Twamley}},\ }\href {https://doi.org/10.1103/PhysRevA.71.012315} {\bibfield
  {journal} {\bibinfo  {journal} {Phys. Rev. A}\ }\textbf {\bibinfo {volume}
  {71}},\ \bibinfo {pages} {012315} (\bibinfo {year} {2005})}\BibitemShut
  {NoStop}%
\bibitem [{\citenamefont {Haljan}\ \emph {et~al.}(2005)\citenamefont {Haljan},
  \citenamefont {Lee}, \citenamefont {Brickman}, \citenamefont {Acton},
  \citenamefont {Deslauriers},\ and\ \citenamefont {Monroe}}]{haljan2005}%
  \BibitemOpen
  \bibfield  {author} {\bibinfo {author} {\bibfnamefont {P.~C.}\ \bibnamefont
  {Haljan}}, \bibinfo {author} {\bibfnamefont {P.~J.}\ \bibnamefont {Lee}},
  \bibinfo {author} {\bibfnamefont {K.-A.}\ \bibnamefont {Brickman}}, \bibinfo
  {author} {\bibfnamefont {M.}~\bibnamefont {Acton}}, \bibinfo {author}
  {\bibfnamefont {L.}~\bibnamefont {Deslauriers}},\ and\ \bibinfo {author}
  {\bibfnamefont {C.}~\bibnamefont {Monroe}},\ }\href
  {https://doi.org/10.1103/PhysRevA.72.062316} {\bibfield  {journal} {\bibinfo
  {journal} {Phys. Rev. A}\ }\textbf {\bibinfo {volume} {72}},\ \bibinfo
  {pages} {062316} (\bibinfo {year} {2005})}\BibitemShut {NoStop}%
\bibitem [{\citenamefont {Chiaverini}\ and\ \citenamefont
  {Lybarger}(2008)}]{chiaverini2008}%
  \BibitemOpen
  \bibfield  {author} {\bibinfo {author} {\bibfnamefont {J.}~\bibnamefont
  {Chiaverini}}\ and\ \bibinfo {author} {\bibfnamefont {W.~E.}\ \bibnamefont
  {Lybarger}},\ }\href {https://doi.org/10.1103/PhysRevA.77.022324} {\bibfield
  {journal} {\bibinfo  {journal} {Phys. Rev. A}\ }\textbf {\bibinfo {volume}
  {77}},\ \bibinfo {pages} {022324} (\bibinfo {year} {2008})}\BibitemShut
  {NoStop}%
\bibitem [{\citenamefont {Johanning}\ \emph {et~al.}(2009)\citenamefont
  {Johanning}, \citenamefont {Braun}, \citenamefont {Timoney}, \citenamefont
  {Elman}, \citenamefont {Neuhauser},\ and\ \citenamefont
  {Wunderlich}}]{johanning2009}%
  \BibitemOpen
  \bibfield  {author} {\bibinfo {author} {\bibfnamefont {M.}~\bibnamefont
  {Johanning}}, \bibinfo {author} {\bibfnamefont {A.}~\bibnamefont {Braun}},
  \bibinfo {author} {\bibfnamefont {N.}~\bibnamefont {Timoney}}, \bibinfo
  {author} {\bibfnamefont {V.}~\bibnamefont {Elman}}, \bibinfo {author}
  {\bibfnamefont {W.}~\bibnamefont {Neuhauser}},\ and\ \bibinfo {author}
  {\bibfnamefont {C.}~\bibnamefont {Wunderlich}},\ }\href
  {https://doi.org/10.1103/PhysRevLett.102.073004} {\bibfield  {journal}
  {\bibinfo  {journal} {Phys. Rev. Lett.}\ }\textbf {\bibinfo {volume} {102}},\
  \bibinfo {pages} {073004} (\bibinfo {year} {2009})}\BibitemShut {NoStop}%
\bibitem [{\citenamefont {Wang}\ \emph {et~al.}(2009)\citenamefont {Wang},
  \citenamefont {Labaziewicz}, \citenamefont {Ge}, \citenamefont {Shewmon},\
  and\ \citenamefont {Chuang}}]{wang2009}%
  \BibitemOpen
  \bibfield  {author} {\bibinfo {author} {\bibfnamefont {S.~X.}\ \bibnamefont
  {Wang}}, \bibinfo {author} {\bibfnamefont {J.}~\bibnamefont {Labaziewicz}},
  \bibinfo {author} {\bibfnamefont {Y.}~\bibnamefont {Ge}}, \bibinfo {author}
  {\bibfnamefont {R.}~\bibnamefont {Shewmon}},\ and\ \bibinfo {author}
  {\bibfnamefont {I.~L.}\ \bibnamefont {Chuang}},\ }\href
  {https://doi.org/10.1063/1.3095520} {\bibfield  {journal} {\bibinfo
  {journal} {Appl. Phys. Lett.}\ }\textbf {\bibinfo {volume} {94}},\ \bibinfo
  {pages} {094103} (\bibinfo {year} {2009})}\BibitemShut {NoStop}%
\bibitem [{\citenamefont {Warring}\ \emph {et~al.}(2013)\citenamefont
  {Warring}, \citenamefont {Ospelkaus}, \citenamefont {Colombe}, \citenamefont
  {J\"ordens}, \citenamefont {Leibfried},\ and\ \citenamefont
  {Wineland}}]{warring_mw_gradient_2013}%
  \BibitemOpen
  \bibfield  {author} {\bibinfo {author} {\bibfnamefont {U.}~\bibnamefont
  {Warring}}, \bibinfo {author} {\bibfnamefont {C.}~\bibnamefont {Ospelkaus}},
  \bibinfo {author} {\bibfnamefont {Y.}~\bibnamefont {Colombe}}, \bibinfo
  {author} {\bibfnamefont {R.}~\bibnamefont {J\"ordens}}, \bibinfo {author}
  {\bibfnamefont {D.}~\bibnamefont {Leibfried}},\ and\ \bibinfo {author}
  {\bibfnamefont {D.~J.}\ \bibnamefont {Wineland}},\ }\href
  {https://doi.org/10.1103/PhysRevLett.110.173002} {\bibfield  {journal}
  {\bibinfo  {journal} {Phys. Rev. Lett.}\ }\textbf {\bibinfo {volume} {110}},\
  \bibinfo {pages} {173002} (\bibinfo {year} {2013})}\BibitemShut {NoStop}%
\bibitem [{\citenamefont {Navon}\ \emph {et~al.}(2013)\citenamefont {Navon},
  \citenamefont {Kotler}, \citenamefont {Akerman}, \citenamefont {Glickman},
  \citenamefont {Almog},\ and\ \citenamefont
  {Ozeri}}]{navon_variable_coupling_2013}%
  \BibitemOpen
  \bibfield  {author} {\bibinfo {author} {\bibfnamefont {N.}~\bibnamefont
  {Navon}}, \bibinfo {author} {\bibfnamefont {S.}~\bibnamefont {Kotler}},
  \bibinfo {author} {\bibfnamefont {N.}~\bibnamefont {Akerman}}, \bibinfo
  {author} {\bibfnamefont {Y.}~\bibnamefont {Glickman}}, \bibinfo {author}
  {\bibfnamefont {I.}~\bibnamefont {Almog}},\ and\ \bibinfo {author}
  {\bibfnamefont {R.}~\bibnamefont {Ozeri}},\ }\href
  {https://doi.org/10.1103/PhysRevLett.111.073001} {\bibfield  {journal}
  {\bibinfo  {journal} {Phys. Rev. Lett.}\ }\textbf {\bibinfo {volume} {111}},\
  \bibinfo {pages} {073001} (\bibinfo {year} {2013})}\BibitemShut {NoStop}%
\bibitem [{\citenamefont {Piltz}\ \emph {et~al.}(2014)\citenamefont {Piltz},
  \citenamefont {Sriarunothai}, \citenamefont {Var{\'o}n},\ and\ \citenamefont
  {Wunderlich}}]{piltz_q_register_2014}%
  \BibitemOpen
  \bibfield  {author} {\bibinfo {author} {\bibfnamefont {C.}~\bibnamefont
  {Piltz}}, \bibinfo {author} {\bibfnamefont {T.}~\bibnamefont {Sriarunothai}},
  \bibinfo {author} {\bibfnamefont {A.~F.}\ \bibnamefont {Var{\'o}n}},\ and\
  \bibinfo {author} {\bibfnamefont {C.}~\bibnamefont {Wunderlich}},\ }\href
  {https://doi.org/10.1038/ncomms5679} {\bibfield  {journal} {\bibinfo
  {journal} {Nat. Comm.}\ }\textbf {\bibinfo {volume} {5}},\ \bibinfo {pages}
  {4679} (\bibinfo {year} {2014})}\BibitemShut {NoStop}%
\bibitem [{\citenamefont {{Aude Craik}}\ \emph {et~al.}(2014)\citenamefont
  {{Aude Craik}}, \citenamefont {{Linke}}, \citenamefont {{Harty}},
  \citenamefont {{Ballance}}, \citenamefont {{Lucas}}, \citenamefont
  {{Steane}},\ and\ \citenamefont
  {{Allcock}}}]{aude-craik_microwave_singlesite_addr_2014}%
  \BibitemOpen
  \bibfield  {author} {\bibinfo {author} {\bibfnamefont {D.~P.~L.}\
  \bibnamefont {{Aude Craik}}}, \bibinfo {author} {\bibfnamefont {N.~M.}\
  \bibnamefont {{Linke}}}, \bibinfo {author} {\bibfnamefont {T.~P.}\
  \bibnamefont {{Harty}}}, \bibinfo {author} {\bibfnamefont {C.~J.}\
  \bibnamefont {{Ballance}}}, \bibinfo {author} {\bibfnamefont {D.~M.}\
  \bibnamefont {{Lucas}}}, \bibinfo {author} {\bibfnamefont {A.~M.}\
  \bibnamefont {{Steane}}},\ and\ \bibinfo {author} {\bibfnamefont {D.~T.~C.}\
  \bibnamefont {{Allcock}}},\ }\href
  {https://doi.org/10.1007/s00340-013-5716-7} {\bibfield  {journal} {\bibinfo
  {journal} {Appl. Phys. B}\ }\textbf {\bibinfo {volume} {114}},\ \bibinfo
  {pages} {3} (\bibinfo {year} {2014})}\BibitemShut {NoStop}%
\bibitem [{\citenamefont {Seck}\ \emph {et~al.}(2020)\citenamefont {Seck},
  \citenamefont {Meier}, \citenamefont {Merrill}, \citenamefont {Hayden},
  \citenamefont {Sawyer}, \citenamefont {Volin},\ and\ \citenamefont
  {Brown}}]{seck_mod_potential_addressing_2020}%
  \BibitemOpen
  \bibfield  {author} {\bibinfo {author} {\bibfnamefont {C.~M.}\ \bibnamefont
  {Seck}}, \bibinfo {author} {\bibfnamefont {A.~M.}\ \bibnamefont {Meier}},
  \bibinfo {author} {\bibfnamefont {J.~T.}\ \bibnamefont {Merrill}}, \bibinfo
  {author} {\bibfnamefont {H.~T.}\ \bibnamefont {Hayden}}, \bibinfo {author}
  {\bibfnamefont {B.~C.}\ \bibnamefont {Sawyer}}, \bibinfo {author}
  {\bibfnamefont {C.~E.}\ \bibnamefont {Volin}},\ and\ \bibinfo {author}
  {\bibfnamefont {K.~R.}\ \bibnamefont {Brown}},\ }\href
  {https://doi.org/10.1088/1367-2630/ab8046} {\bibfield  {journal} {\bibinfo
  {journal} {New J. Phys.}\ }\textbf {\bibinfo {volume} {22}},\ \bibinfo
  {pages} {053024} (\bibinfo {year} {2020})}\BibitemShut {NoStop}%
\bibitem [{\citenamefont {Srinivas}\ \emph {et~al.}(2021)\citenamefont
  {Srinivas}, \citenamefont {Burd}, \citenamefont {Knaack}, \citenamefont
  {Sutherland}, \citenamefont {Kwiatkowski}, \citenamefont {Glancy},
  \citenamefont {Knill}, \citenamefont {Wineland}, \citenamefont {Leibfried},
  \citenamefont {Wilson}, \citenamefont {Allcock},\ and\ \citenamefont
  {Slichter}}]{srinivas_high_fid_laser_free_2021}%
  \BibitemOpen
  \bibfield  {author} {\bibinfo {author} {\bibfnamefont {R.}~\bibnamefont
  {Srinivas}}, \bibinfo {author} {\bibfnamefont {S.~C.}\ \bibnamefont {Burd}},
  \bibinfo {author} {\bibfnamefont {H.~M.}\ \bibnamefont {Knaack}}, \bibinfo
  {author} {\bibfnamefont {R.~T.}\ \bibnamefont {Sutherland}}, \bibinfo
  {author} {\bibfnamefont {A.}~\bibnamefont {Kwiatkowski}}, \bibinfo {author}
  {\bibfnamefont {S.}~\bibnamefont {Glancy}}, \bibinfo {author} {\bibfnamefont
  {E.}~\bibnamefont {Knill}}, \bibinfo {author} {\bibfnamefont {D.~J.}\
  \bibnamefont {Wineland}}, \bibinfo {author} {\bibfnamefont {D.}~\bibnamefont
  {Leibfried}}, \bibinfo {author} {\bibfnamefont {A.~C.}\ \bibnamefont
  {Wilson}}, \bibinfo {author} {\bibfnamefont {D.~T.~C.}\ \bibnamefont
  {Allcock}},\ and\ \bibinfo {author} {\bibfnamefont {D.~H.}\ \bibnamefont
  {Slichter}},\ }\href {https://doi.org/10.1038/s41586-021-03809-4} {\bibfield
  {journal} {\bibinfo  {journal} {Nature}\ }\textbf {\bibinfo {volume} {597}},\
  \bibinfo {pages} {209} (\bibinfo {year} {2021})}\BibitemShut {NoStop}%
\bibitem [{\citenamefont {Sutherland}\ \emph {et~al.}(2023)\citenamefont
  {Sutherland}, \citenamefont {Srinivas},\ and\ \citenamefont
  {Allcock}}]{sutherland_geo_phase_addressing_2023}%
  \BibitemOpen
  \bibfield  {author} {\bibinfo {author} {\bibfnamefont {R.~T.}\ \bibnamefont
  {Sutherland}}, \bibinfo {author} {\bibfnamefont {R.}~\bibnamefont
  {Srinivas}},\ and\ \bibinfo {author} {\bibfnamefont {D.~T.~C.}\ \bibnamefont
  {Allcock}},\ }\href {https://doi.org/10.1103/PhysRevA.107.032604} {\bibfield
  {journal} {\bibinfo  {journal} {Phys. Rev. A}\ }\textbf {\bibinfo {volume}
  {107}},\ \bibinfo {pages} {032604} (\bibinfo {year} {2023})}\BibitemShut
  {NoStop}%
\bibitem [{\citenamefont {Srinivas}\ \emph {et~al.}(2023)\citenamefont
  {Srinivas}, \citenamefont {L\"oschnauer}, \citenamefont {Malinowski},
  \citenamefont {Hughes}, \citenamefont {Nourshargh}, \citenamefont
  {Negnevitsky}, \citenamefont {Allcock}, \citenamefont {King}, \citenamefont
  {Matthiesen}, \citenamefont {Harty},\ and\ \citenamefont
  {Ballance}}]{srinivas_local_fields_2023}%
  \BibitemOpen
  \bibfield  {author} {\bibinfo {author} {\bibfnamefont {R.}~\bibnamefont
  {Srinivas}}, \bibinfo {author} {\bibfnamefont {C.~M.}\ \bibnamefont
  {L\"oschnauer}}, \bibinfo {author} {\bibfnamefont {M.}~\bibnamefont
  {Malinowski}}, \bibinfo {author} {\bibfnamefont {A.~C.}\ \bibnamefont
  {Hughes}}, \bibinfo {author} {\bibfnamefont {R.}~\bibnamefont {Nourshargh}},
  \bibinfo {author} {\bibfnamefont {V.}~\bibnamefont {Negnevitsky}}, \bibinfo
  {author} {\bibfnamefont {D.~T.~C.}\ \bibnamefont {Allcock}}, \bibinfo
  {author} {\bibfnamefont {S.~A.}\ \bibnamefont {King}}, \bibinfo {author}
  {\bibfnamefont {C.}~\bibnamefont {Matthiesen}}, \bibinfo {author}
  {\bibfnamefont {T.~P.}\ \bibnamefont {Harty}},\ and\ \bibinfo {author}
  {\bibfnamefont {C.~J.}\ \bibnamefont {Ballance}},\ }\href
  {https://doi.org/10.1103/PhysRevLett.131.020601} {\bibfield  {journal}
  {\bibinfo  {journal} {Phys. Rev. Lett.}\ }\textbf {\bibinfo {volume} {131}},\
  \bibinfo {pages} {020601} (\bibinfo {year} {2023})}\BibitemShut {NoStop}%
\bibitem [{\citenamefont {Berkeland}\ \emph {et~al.}(1998)\citenamefont
  {Berkeland}, \citenamefont {Miller}, \citenamefont {Bergquist}, \citenamefont
  {Itano},\ and\ \citenamefont {Wineland}}]{berkeland_minimization_1998}%
  \BibitemOpen
  \bibfield  {author} {\bibinfo {author} {\bibfnamefont {D.~J.}\ \bibnamefont
  {Berkeland}}, \bibinfo {author} {\bibfnamefont {J.~D.}\ \bibnamefont
  {Miller}}, \bibinfo {author} {\bibfnamefont {J.~C.}\ \bibnamefont
  {Bergquist}}, \bibinfo {author} {\bibfnamefont {W.~M.}\ \bibnamefont
  {Itano}},\ and\ \bibinfo {author} {\bibfnamefont {D.~J.}\ \bibnamefont
  {Wineland}},\ }\href {https://doi.org/10.1063/1.367318} {\bibfield  {journal}
  {\bibinfo  {journal} {J. App. Phys.}\ }\textbf {\bibinfo {volume} {83}},\
  \bibinfo {pages} {5025} (\bibinfo {year} {1998})}\BibitemShut {NoStop}%
\bibitem [{\citenamefont {Keller}\ \emph {et~al.}(2015)\citenamefont {Keller},
  \citenamefont {Partner}, \citenamefont {Burgermeister},\ and\ \citenamefont
  {Mehlstaeubler}}]{keller_precise_2015}%
  \BibitemOpen
  \bibfield  {author} {\bibinfo {author} {\bibfnamefont {J.}~\bibnamefont
  {Keller}}, \bibinfo {author} {\bibfnamefont {H.~L.}\ \bibnamefont {Partner}},
  \bibinfo {author} {\bibfnamefont {T.}~\bibnamefont {Burgermeister}},\ and\
  \bibinfo {author} {\bibfnamefont {T.~E.}\ \bibnamefont {Mehlstaeubler}},\
  }\href {https://doi.org/10.1063/1.4930037} {\bibfield  {journal} {\bibinfo
  {journal} {J. App. Phys.}\ }\textbf {\bibinfo {volume} {118}},\ \bibinfo
  {pages} {104501} (\bibinfo {year} {2015})}\BibitemShut {NoStop}%
\bibitem [{\citenamefont {Nadlinger}\ \emph {et~al.}(2021)\citenamefont
  {Nadlinger}, \citenamefont {Drmota}, \citenamefont {Main}, \citenamefont
  {Nichol}, \citenamefont {Araneda}, \citenamefont {Srinivas}, \citenamefont
  {Stephenson}, \citenamefont {Ballance},\ and\ \citenamefont
  {Lucas}}]{nadlinger_micromotion_2021}%
  \BibitemOpen
  \bibfield  {author} {\bibinfo {author} {\bibfnamefont {D.~P.}\ \bibnamefont
  {Nadlinger}}, \bibinfo {author} {\bibfnamefont {P.}~\bibnamefont {Drmota}},
  \bibinfo {author} {\bibfnamefont {D.}~\bibnamefont {Main}}, \bibinfo {author}
  {\bibfnamefont {B.~C.}\ \bibnamefont {Nichol}}, \bibinfo {author}
  {\bibfnamefont {G.}~\bibnamefont {Araneda}}, \bibinfo {author} {\bibfnamefont
  {R.}~\bibnamefont {Srinivas}}, \bibinfo {author} {\bibfnamefont {L.~J.}\
  \bibnamefont {Stephenson}}, \bibinfo {author} {\bibfnamefont {C.~J.}\
  \bibnamefont {Ballance}},\ and\ \bibinfo {author} {\bibfnamefont {D.~M.}\
  \bibnamefont {Lucas}},\ }\Eprint {https://arxiv.org/abs/2107.00056}
  {arXiv:2107.00056 [quant-ph]}  (\bibinfo {year} {2021})\BibitemShut {NoStop}%
\bibitem [{\citenamefont {Bermudez}\ \emph {et~al.}(2017)\citenamefont
  {Bermudez}, \citenamefont {Schindler}, \citenamefont {Monz}, \citenamefont
  {Blatt},\ and\ \citenamefont {Müller}}]{bermudez_micromotion-enabled_2017}%
  \BibitemOpen
  \bibfield  {author} {\bibinfo {author} {\bibfnamefont {A.}~\bibnamefont
  {Bermudez}}, \bibinfo {author} {\bibfnamefont {P.}~\bibnamefont {Schindler}},
  \bibinfo {author} {\bibfnamefont {T.}~\bibnamefont {Monz}}, \bibinfo {author}
  {\bibfnamefont {R.}~\bibnamefont {Blatt}},\ and\ \bibinfo {author}
  {\bibfnamefont {M.}~\bibnamefont {Müller}},\ }\href
  {https://doi.org/10.1088/1367-2630/aa86eb} {\bibfield  {journal} {\bibinfo
  {journal} {New J. Phys.}\ }\textbf {\bibinfo {volume} {19}},\ \bibinfo
  {pages} {113038} (\bibinfo {year} {2017})}\BibitemShut {NoStop}%
\bibitem [{\citenamefont {Gaebler}\ \emph {et~al.}(2021)\citenamefont
  {Gaebler}, \citenamefont {Baldwin}, \citenamefont {Moses}, \citenamefont
  {Dreiling}, \citenamefont {Figgatt}, \citenamefont {Foss-Feig}, \citenamefont
  {Hayes},\ and\ \citenamefont {Pino}}]{gaebler_suppression_2021}%
  \BibitemOpen
  \bibfield  {author} {\bibinfo {author} {\bibfnamefont {J.~P.}\ \bibnamefont
  {Gaebler}}, \bibinfo {author} {\bibfnamefont {C.~H.}\ \bibnamefont
  {Baldwin}}, \bibinfo {author} {\bibfnamefont {S.~A.}\ \bibnamefont {Moses}},
  \bibinfo {author} {\bibfnamefont {J.~M.}\ \bibnamefont {Dreiling}}, \bibinfo
  {author} {\bibfnamefont {C.}~\bibnamefont {Figgatt}}, \bibinfo {author}
  {\bibfnamefont {M.}~\bibnamefont {Foss-Feig}}, \bibinfo {author}
  {\bibfnamefont {D.}~\bibnamefont {Hayes}},\ and\ \bibinfo {author}
  {\bibfnamefont {J.~M.}\ \bibnamefont {Pino}},\ }\href
  {https://doi.org/10.1103/PhysRevA.104.062440} {\bibfield  {journal} {\bibinfo
   {journal} {Phys. Rev. A}\ }\textbf {\bibinfo {volume} {104}},\ \bibinfo
  {pages} {062440} (\bibinfo {year} {2021})}\BibitemShut {NoStop}%
\bibitem [{\citenamefont {Wineland}\ and\ \citenamefont
  {Itano}(1979)}]{wineland_laser_cooling_1979}%
  \BibitemOpen
  \bibfield  {author} {\bibinfo {author} {\bibfnamefont {D.~J.}\ \bibnamefont
  {Wineland}}\ and\ \bibinfo {author} {\bibfnamefont {W.~M.}\ \bibnamefont
  {Itano}},\ }\href {https://doi.org/10.1103/PhysRevA.20.1521} {\bibfield
  {journal} {\bibinfo  {journal} {Phys. Rev. A}\ }\textbf {\bibinfo {volume}
  {20}},\ \bibinfo {pages} {1521} (\bibinfo {year} {1979})}\BibitemShut
  {NoStop}%
\bibitem [{\citenamefont {Dehmelt}(1968)}]{dehmelt_rf_spectra_i_1968}%
  \BibitemOpen
  \bibfield  {author} {\bibinfo {author} {\bibfnamefont {H.~G.}\ \bibnamefont
  {Dehmelt}},\ }in\ \href {https://doi.org/10.1016/S0065-2199(08)60170-0}
  {\emph {\bibinfo {booktitle} {Advances in {{Atomic}} and {{Molecular
  Physics}}}}},\ Vol.~\bibinfo {volume} {3},\ \bibinfo {editor} {edited by\
  \bibinfo {editor} {\bibfnamefont {D.~R.}\ \bibnamefont {Bates}}\ and\
  \bibinfo {editor} {\bibfnamefont {I.}~\bibnamefont {Estermann}}}\ (\bibinfo
  {publisher} {{Academic Press}},\ \bibinfo {year} {1968})\ pp.\ \bibinfo
  {pages} {53--72}\BibitemShut {NoStop}%
\bibitem [{\citenamefont {Douglas}\ \emph {et~al.}(2015)\citenamefont
  {Douglas}, \citenamefont {Berdnikov},\ and\ \citenamefont
  {Konenkov}}]{douglas_effective_rf_motion_2015}%
  \BibitemOpen
  \bibfield  {author} {\bibinfo {author} {\bibfnamefont {D.~J.}\ \bibnamefont
  {Douglas}}, \bibinfo {author} {\bibfnamefont {A.~S.}\ \bibnamefont
  {Berdnikov}},\ and\ \bibinfo {author} {\bibfnamefont {N.~V.}\ \bibnamefont
  {Konenkov}},\ }\href {https://doi.org/10.1016/j.ijms.2014.08.009} {\bibfield
  {journal} {\bibinfo  {journal} {Int. J. Mass Spectrom.}\ }\textbf {\bibinfo
  {volume} {377}},\ \bibinfo {pages} {345} (\bibinfo {year}
  {2015})}\BibitemShut {NoStop}%
\bibitem [{\citenamefont {Wesenberg}(2008)}]{wesenberg_electrostatics_2008}%
  \BibitemOpen
  \bibfield  {author} {\bibinfo {author} {\bibfnamefont {J.~H.}\ \bibnamefont
  {Wesenberg}},\ }\href {https://doi.org/10.1103/PhysRevA.78.063410} {\bibfield
   {journal} {\bibinfo  {journal} {Phys. Rev. A}\ }\textbf {\bibinfo {volume}
  {78}},\ \bibinfo {pages} {063410} (\bibinfo {year} {2008})}\BibitemShut
  {NoStop}%
\bibitem [{\citenamefont {Kwon}\ \emph {et~al.}(2023)\citenamefont {Kwon},
  \citenamefont {Setzer}, \citenamefont {Gehl}, \citenamefont {Karl},
  \citenamefont {Wall}, \citenamefont {Law}, \citenamefont {Stick},\ and\
  \citenamefont {McGuinness}}]{kwon_multisite_optical_2023}%
  \BibitemOpen
  \bibfield  {author} {\bibinfo {author} {\bibfnamefont {J.}~\bibnamefont
  {Kwon}}, \bibinfo {author} {\bibfnamefont {W.~J.}\ \bibnamefont {Setzer}},
  \bibinfo {author} {\bibfnamefont {M.}~\bibnamefont {Gehl}}, \bibinfo {author}
  {\bibfnamefont {N.}~\bibnamefont {Karl}}, \bibinfo {author} {\bibfnamefont
  {J.~V.~D.}\ \bibnamefont {Wall}}, \bibinfo {author} {\bibfnamefont
  {R.}~\bibnamefont {Law}}, \bibinfo {author} {\bibfnamefont {D.}~\bibnamefont
  {Stick}},\ and\ \bibinfo {author} {\bibfnamefont {H.~J.}\ \bibnamefont
  {McGuinness}},\ }\Eprint {https://arxiv.org/abs/2308.14918} {arXiv:2308.14918
  [quant-ph]}  (\bibinfo {year} {2023})\BibitemShut {NoStop}%
\bibitem [{\citenamefont {Mehta}\ \emph {et~al.}(2016)\citenamefont {Mehta},
  \citenamefont {Bruzewicz}, \citenamefont {McConnell}, \citenamefont {Ram},
  \citenamefont {Sage},\ and\ \citenamefont
  {Chiaverini}}]{mehta_inegrated_optics_2016}%
  \BibitemOpen
  \bibfield  {author} {\bibinfo {author} {\bibfnamefont {K.~K.}\ \bibnamefont
  {Mehta}}, \bibinfo {author} {\bibfnamefont {C.~D.}\ \bibnamefont
  {Bruzewicz}}, \bibinfo {author} {\bibfnamefont {R.}~\bibnamefont
  {McConnell}}, \bibinfo {author} {\bibfnamefont {R.~J.}\ \bibnamefont {Ram}},
  \bibinfo {author} {\bibfnamefont {J.~M.}\ \bibnamefont {Sage}},\ and\
  \bibinfo {author} {\bibfnamefont {J.}~\bibnamefont {Chiaverini}},\ }\href
  {https://doi.org/10.1038/nnano.2016.139} {\bibfield  {journal} {\bibinfo
  {journal} {Nat. Nano.}\ }\textbf {\bibinfo {volume} {11}},\ \bibinfo {pages}
  {1066} (\bibinfo {year} {2016})}\BibitemShut {NoStop}%
\bibitem [{\citenamefont {Mehta}\ \emph {et~al.}(2020)\citenamefont {Mehta},
  \citenamefont {Zhang}, \citenamefont {Malinowski}, \citenamefont {Nguyen},
  \citenamefont {Stadler},\ and\ \citenamefont {Home}}]{mehta_multiion_2020}%
  \BibitemOpen
  \bibfield  {author} {\bibinfo {author} {\bibfnamefont {K.~K.}\ \bibnamefont
  {Mehta}}, \bibinfo {author} {\bibfnamefont {C.}~\bibnamefont {Zhang}},
  \bibinfo {author} {\bibfnamefont {M.}~\bibnamefont {Malinowski}}, \bibinfo
  {author} {\bibfnamefont {T.-L.}\ \bibnamefont {Nguyen}}, \bibinfo {author}
  {\bibfnamefont {M.}~\bibnamefont {Stadler}},\ and\ \bibinfo {author}
  {\bibfnamefont {J.~P.}\ \bibnamefont {Home}},\ }\href
  {https://doi.org/10.1038/s41586-020-2823-6} {\bibfield  {journal} {\bibinfo
  {journal} {Nature}\ }\textbf {\bibinfo {volume} {586}},\ \bibinfo {pages}
  {533} (\bibinfo {year} {2020})}\BibitemShut {NoStop}%
\bibitem [{\citenamefont {Niffenegger}\ \emph {et~al.}(2020)\citenamefont
  {Niffenegger}, \citenamefont {Stuart}, \citenamefont {Sorace-Agaskar},
  \citenamefont {Kharas}, \citenamefont {Bramhavar}, \citenamefont {Bruzewicz},
  \citenamefont {Loh}, \citenamefont {Maxson}, \citenamefont {McConnell},
  \citenamefont {Reens}, \citenamefont {West}, \citenamefont {Sage},\ and\
  \citenamefont {Chiaverini}}]{niffenegger_multiwav_2020}%
  \BibitemOpen
  \bibfield  {author} {\bibinfo {author} {\bibfnamefont {R.~J.}\ \bibnamefont
  {Niffenegger}}, \bibinfo {author} {\bibfnamefont {J.}~\bibnamefont {Stuart}},
  \bibinfo {author} {\bibfnamefont {C.}~\bibnamefont {Sorace-Agaskar}},
  \bibinfo {author} {\bibfnamefont {D.}~\bibnamefont {Kharas}}, \bibinfo
  {author} {\bibfnamefont {S.}~\bibnamefont {Bramhavar}}, \bibinfo {author}
  {\bibfnamefont {C.~D.}\ \bibnamefont {Bruzewicz}}, \bibinfo {author}
  {\bibfnamefont {W.}~\bibnamefont {Loh}}, \bibinfo {author} {\bibfnamefont
  {R.~T.}\ \bibnamefont {Maxson}}, \bibinfo {author} {\bibfnamefont
  {R.}~\bibnamefont {McConnell}}, \bibinfo {author} {\bibfnamefont
  {D.}~\bibnamefont {Reens}}, \bibinfo {author} {\bibfnamefont {G.~N.}\
  \bibnamefont {West}}, \bibinfo {author} {\bibfnamefont {J.~M.}\ \bibnamefont
  {Sage}},\ and\ \bibinfo {author} {\bibfnamefont {J.}~\bibnamefont
  {Chiaverini}},\ }\href {https://doi.org/10.1038/s41586-020-2811-x} {\bibfield
   {journal} {\bibinfo  {journal} {Nature}\ }\textbf {\bibinfo {volume}
  {586}},\ \bibinfo {pages} {538} (\bibinfo {year} {2020})}\BibitemShut
  {NoStop}%
\end{thebibliography}%
\bibliographystyle{apsrev4-2}

\end{document}